\documentclass[lettersize,journal]{IEEEtran}
\usepackage{algorithmic}
\usepackage{algorithm}
\usepackage{array}
\usepackage[caption=false,font=normalsize,labelfont=sf,textfont=sf]{subfig}
\usepackage{textcomp}
\usepackage{stfloats}
\usepackage{url}
\usepackage{verbatim}
\usepackage{graphicx}
\usepackage{cite}
\usepackage{booktabs}
\usepackage{amsthm}          
\usepackage{amsmath, amssymb, amsfonts} 
\newcommand{\norm}[1]{\left\| #1 \right\|} 
  
\usepackage{color}

\begin{document}

\title{Joint Channel Estimation and Dynamics-Aware Grouping for Time-Varying RIS-Assisted OTA Federated Learning}

\author{Ziqi Li, Shuangzhi Li, Uchechukwu Awada, and Jiankang Zhang%
\thanks{Ziqi Li is with the School of Electrical and Information Engineering, Zhengzhou University, Zhengzhou 450001, China, and also with the Department of Electrical and Computer Engineering, The University of Hong Kong, Hong Kong SAR, China ({Email:ziqili0901@connect.hku.hk}).}%
\thanks{Shuangzhi Li is with the School of Electrical and Information Engineering, Zhengzhou University, Zhengzhou 450001, China (Email:ielsz@zzu.edu.cn).}%
\thanks{Uchechukwu Awada is with the School of Software, Henan Institute of Science and Technology,
Xinxiang, 453003, China (E-mail: awada@hist.edu.cn).} 
\thanks{Jiankang Zhang is with the School of Computing and Engineering, Bournemouth University, UK (Email: jzhang3@bournemouth.ac.uk)}
\thanks{Shuangzhi Li is the corresponding author (Email: ielsz@zzu.edu.cn).}}

\maketitle

\begin{abstract}
Reconfigurable intelligent surface (RIS)-assisted over-the-air federated learning (OTA-FL) enables efficient distributed intelligence but suffers from time-varying channels, imperfect channel state information (CSI), and strong user heterogeneity, which jointly degrade aggregation accuracy and cause severe model update cancellation. To address these issues, we propose a unified framework for joint channel estimation and dynamics-aware user grouping in RIS-assisted OTA-FL systems, enabling reliable learning under imperfect CSI and heterogeneous dynamics. The framework integrates gated recurrent unit (GRU) for temporal modeling to capture time-varying CSI evolution, OTA-based federated aggregation with personalization, and RIS-aware physical-layer optimization in a closed loop. In addition, we design a dynamics-aware grouping strategy based on long-term path-loss and short-term channel dynamics to reduce inter-user conflicts under heterogeneous conditions. Simulation results show that the proposed method achieves substantial gains in CSI estimation accuracy and OTA aggregation performance in low-pilot and high-mobility regimes, while improving convergence speed and robustness under strong user heterogeneity.
\end{abstract}


\section{Introduction}
As 6G wireless networks evolve toward edge intelligence, federated learning (FL) has emerged as a promising paradigm for distributed model training without sharing raw data. By enabling local training at devices and transmitting only model updates, FL significantly reduces communication overhead and preserves data privacy compared with conventional centralized learning \cite{8016573,8030322,9599369,9625822, 2016arXiv160205629M, 2019arXiv191204977K}. However, its iterative training process introduces substantial uplink communication bottlenecks, especially in large-scale wireless networks \cite{2016arXiv161002527K}.

To address this limitation, over-the-air federated learning (OTA-FL) has been proposed by exploiting the superposition property of wireless channels, allowing simultaneous transmission of model updates and direct aggregation at the base station \cite{8870236,9042352, 9611013}. Despite its communication efficiency, OTA-FL is highly sensitive to channel state information (CSI) accuracy. In practice, imperfect CSI inevitably induces aggregation distortion, which in turn degrades learning performance \cite{8952884,10261509}. This issue is further exacerbated in time-varying wireless environments, where channel dynamics fundamentally complicate reliable aggregation.

Reconfigurable intelligent surfaces (RIS) provide a promising solution by reshaping wireless propagation environments through controllable phase shifts, thereby improving link quality and mitigating aggregation errors \cite{8811733,9013643,9199786,11231091,LI2024109611}. \cite{10145869} derived an upper bound on the aggregation error of RIS-assisted OTA systems, and subsequent studies further verified the tangible performance gains brought by RIS to over-the-air aggregation \cite{9013643,9199786}. Nevertheless, RIS-assisted OTA-FL introduces tightly coupled challenges, including cascaded channel estimation, pilot overhead limitation, and time-varying channel dynamics \cite{9771077,9722893,10.1093/nsr/nwad127}. Moreover, heterogeneous channel conditions across users lead to so-called tail devices, whose poor link quality and dynamic behavior significantly degrade aggregation reliability while remaining essential for maintaining data diversity and fairness \cite{9400843}.

To address these challenges, recent studies have explored device selection, clustered federated learning, and meta-learning-based personalization. For instance, \cite{8870236} excluded straggling users from model aggregation to reduce communication errors, while \cite{9451567} jointly considered device selection and communication error, developing a Gibbs-sampling-based strategy to improve overall system performance. However, existing device selection methods often sacrifice weak users to improve learning efficiency. While this design is commonly adopted in application-layer federated learning, it may introduce fairness concerns in wireless CSI acquisition, where reliable channel information should be collected across diverse propagation conditions rather than only from strong-link users. Moreover, retaining weak users can provide additional benefits for channel modeling, as their CSI captures rich propagation diversity under low-SNR and non-line-of-sight conditions. This diversity can potentially enhance the robustness of channel estimation models under dynamic wireless environments. Clustering methods also achieved more stable region-specific channel modeling\cite{10884956,Long2023MultiCenterFL,9174890,9832954}, but suffer from high overhead and poor adaptability in dynamic wireless environments, due to their static and initialization-sensitive design. Meanwhile, personalized FL methods, including multi-task federated learning \cite{NIPS2017_6211080f} and meta-learning approaches such as Per-FedAvg \cite{2020arXiv200207948F}, FedMeta \cite{2018arXiv180207876C} and pFedMe \cite{2020arXiv200608848D}, improve personalization in federated learning. However, these methods, along with most existing FL optimization frameworks, commonly assume ideal or static wireless channels and thus overlook the coupling between time-varying channel dynamics and federated optimization. This mismatch limits their applicability in wireless-enabled FL systems, where communication reliability and learning performance are jointly affected by channel dynamics.
Despite these advances, a fundamental challenge remains: in RIS-assisted OTA-FL systems, channel estimation, RIS configuration, and federated learning optimization are tightly coupled under time-varying wireless conditions, making separate or sequential design fundamentally suboptimal. This coupling creates a critical gap between communication reliability and learning stability, which cannot be effectively bridged by existing decoupled approaches.

Motivated by this observation, we study time-varying RIS-assisted OTA-FL under imperfect cascaded CSI, and propose a unified framework that jointly addresses channel estimation, communication distortion, and user heterogeneity. The framework integrates GRU-based temporal modeling for CSI prediction, personalized OTA-FL aggregation, and RIS-aware physical-layer optimization in a closed-loop design. Furthermore, we introduce a dynamics-aware grouping strategy to adaptively mitigate heterogeneity across users. The main contributions of this paper are summarized as follows:
\begin{itemize}
    \item We propose a system-level closed-loop RIS-assisted OTA-FL framework that jointly optimizes channel estimation, RIS phase control, and federated aggregation under imperfect CSI, explicitly incorporating OTA-induced distortion into the learning process.

    \item We design a personalized federated learning architecture that employs gated recurrent units (GRUs) as the temporal backbone, enabling accurate prediction of time-varying cascaded CSI while supporting heterogeneous local adaptation.

    \item We develop a dynamics-aware grouping strategy that exploits long-term path-loss and short-term channel dynamics to improve aggregation stability under strong user heterogeneity.
\end{itemize}

We use $\mathbb{R}$ and $\mathbb{C}$ to denote the sets of real and complex numbers, respectively. Lowercase letters, bold lowercase letters, and bold uppercase letters are used to denote scalars, vectors, and matrices, respectively. Calligraphic letters are used to denote sets, groups, filtrations, or operators, depending on the context. $\mathcal{L}$ denotes the loss function. The cardinality of a set $\mathcal{D}$ is denoted by $|\mathcal{D}|$. $r$ denotes the time index corresponding to the $r$-th federated learning round, and $k \in {1,\dots,K}$ denotes the user index. The operators $(\cdot)^T$, $(\cdot)^H$, $(\cdot)^*$, $\odot$, $\Re{\cdot}$, $\Im{\cdot}$, $[\cdot;\cdot]$, and $\mathbb{I}(\cdot)$ denote the transpose, Hermitian transpose, complex conjugate, Hadamard product, real part, imaginary part, vertical concatenation, and indicator function, respectively, where $\mathbb{I}(\cdot)$ equals one if the enclosed condition holds and zero otherwise. $\operatorname{diag}(\mathbf{x})$ denotes a diagonal matrix whose diagonal entries are given by $\mathbf{x}$. We use $|\cdot|_2$ to denote the Euclidean norm, $\mathbb{E}[\cdot]$ to denote the expectation operator, $\mathbf{I}_N$ to denote the $N \times N$ identity matrix, $\mathcal{CN}(\mu,\sigma^2)$ for a scalar circularly symmetric complex Gaussian distribution with mean $\mu$ and variance $\sigma^2$, and $\mathcal{CN}(\boldsymbol{\mu},\mathbf{\Sigma})$ for its vector counterpart with mean vector $\boldsymbol{\mu}$ and covariance matrix $\mathbf{\Sigma}$.

\section{System model}

\subsection{Federated Learning System}\label{sec:FL}
We consider a wireless federated learning (FL) system consisting of one multi-antenna base station (BS) that also serves as the parameter server, one reconfigurable intelligent surface (RIS), and $K$ single-antenna edge users. The BS broadcasts the global model and aggregates the local model updates uploaded by the users, while each user performs local training using its own data. The RIS is deployed to assist the underlying wireless transmission.

Let the global model parameter (or its shared component) be $\mathbf{w} \in \mathbb{R}^{D}$, where $D$ denotes the dimension of the model parameters. User $k$ holds a local dataset $\mathcal{D}_k = \left\{ (\mathbf{x}_{k,i}, \mathbf{z}_{k,i}): 1 \leq i \leq Q_k \right\}$, where $Q_k = |\mathcal{D}_k|$ is the number of local samples of user $k$, $\mathbf{x}_{k,i}$ denotes the input feature, and $\mathbf{z}_{k,i}$ denotes the corresponding label. The total number of samples across all users is $Q = \sum_{k=1}^{K} Q_k$.

The overall FL task can be formulated as the following empirical risk minimization problem:
\begin{equation}
\min_{\mathbf{w} \in \mathbb{R}^{D}} F(\mathbf{w}) = \frac{1}{Q} \sum_{k=1}^{K} Q_k F_k(\mathbf{w}; \mathcal{D}_k),
\end{equation}
where
$$F_k(\mathbf{w}; \mathcal{D}_k) \triangleq \frac{1}{Q_k} \sum_{(\mathbf{x}_{k,i}, \mathbf{z}_{k,i}) \in \mathcal{D}_k} \ell(\mathbf{w}; \mathbf{x}_{k,i}, \mathbf{z}_{k,i})$$ is the local empirical loss function of user $k$, and $\ell(\cdot)$ denotes the per-sample loss function.

The FL training procedure consists of $R$ iterative rounds. In the $r$-th round, the BS first broadcasts the current global model $\mathbf{w}^{r}$ to all users. Each user $k$ then performs local training on its local data and obtains the updated local model:
\begin{equation}
\mathbf{w}_{k,\mathrm{loc}}^{r} = \mathcal{U}_k(\mathbf{w}^{r}; \mathcal{D}_k^{r}),
\end{equation}
where $\mathcal{U}_k(\cdot)$ denotes the local training operator of user $k$, and $\mathcal{D}_k^{r}$ is the set of local samples used in the $r$-th round. After local training, user $k$ computes the model update
\begin{equation}
\Delta \mathbf{w}_k^{r} = \mathbf{w}_{k,\mathrm{loc}}^{r} - \mathbf{w}^{r}.
\end{equation}

Since the number of local samples may vary across users, model aggregation is typically weighted by the sample size. We define the user weight in the $r$-th round as $\omega_k^{r} = \frac{Q_k^{r}}{\frac{1}{K}\sum_{i=1}^{K} Q_i^{r}}$, where $Q_k^{r} = |\mathcal{D}_k^{r}|$ is the number of local samples used by user $k$ in the $r$-th round. The global weighted update that the BS expects to recover under ideal conditions is
\begin{equation}
\mathbf{r}^{r} \triangleq \sum_{k=1}^{K} \omega_k^{r} \Delta \mathbf{w}_k^{r}.
\end{equation}

Let the estimated value of the global update recovered at the receiver be $\hat{\mathbf{r}}^{r}$. The global model update is then updated as
\begin{equation}
\mathbf{w}^{r+1} = \mathbf{w}^{r} + \frac{1}{\sum_{k=1}^{K} \omega_k^{r}} \hat{\mathbf{r}}^{r}.
\end{equation}

In summary, the FL process in this system follows a standard closed-loop procedure: model broadcasting, local training, wireless aggregation, and global update. The underlying wireless channel only affects the recovery accuracy of $\hat{\mathbf{r}}^{r}$ with respect to $\mathbf{r}^{r}$.

\subsection{RIS-Assisted Pilot Observation Model}
\label{sec:Pilot}
\begin{figure}
\centering
\includegraphics[width=\linewidth]{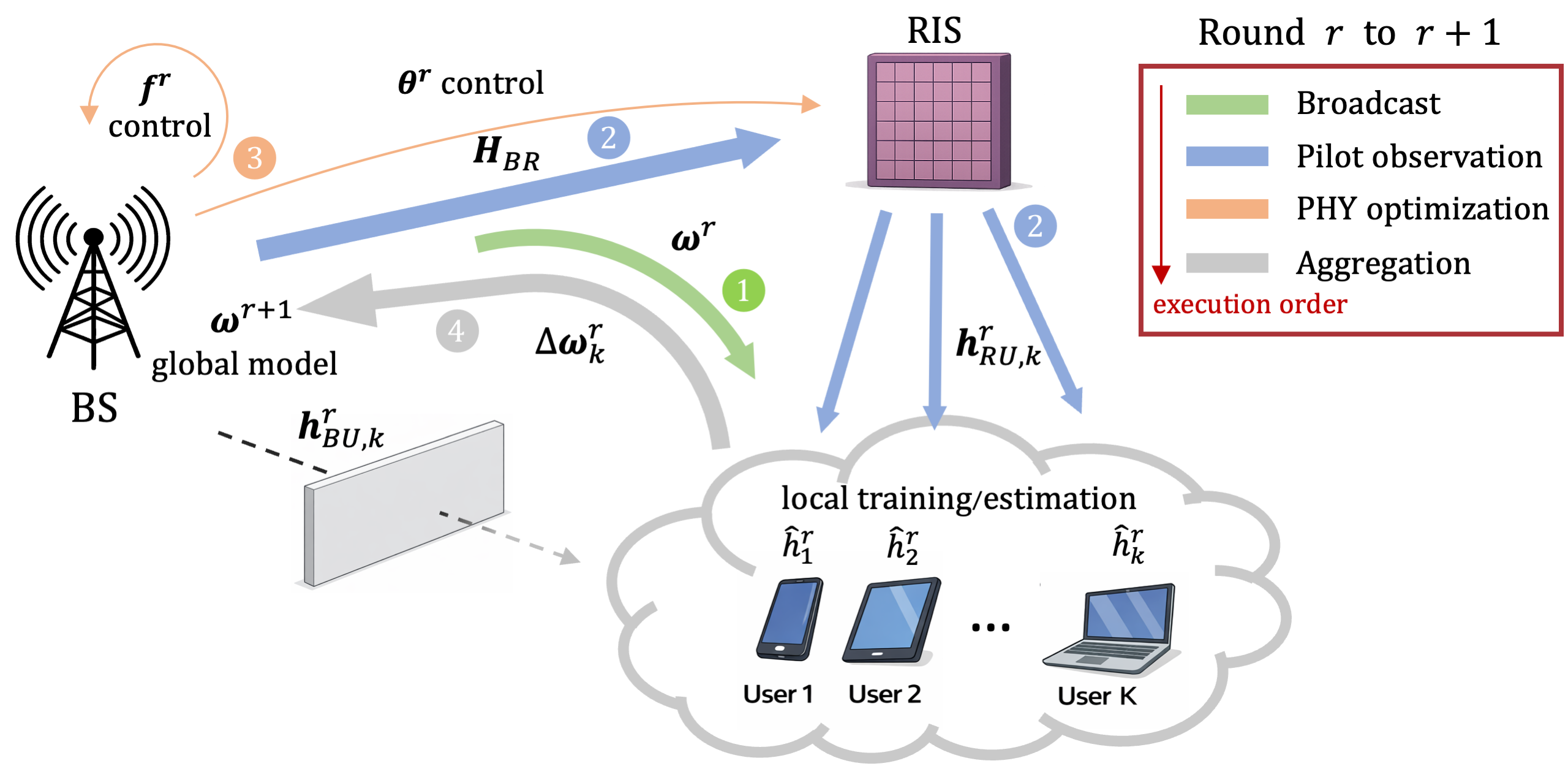}
\caption{RIS-assisted cascaded-channel OTA-FL system model and per-round procedure, where PHY denotes the physical layer (detailed in Section~\ref{sec:Round}). The direct link is assumed blocked or weak.}
\label{fig:system}
\end{figure}

The FL system described above operates in an RIS-assisted wireless environment. The BS is equipped with $M$ antennas, the RIS consists of $N$ reflecting elements, and all users are single-antenna devices. Let the BS-to-RIS channel matrix be $\mathbf{H}_{BR} \in \mathbb{C}^{N \times M}$, the RIS-to-user channel for user $k$ be $\mathbf{h}_{RU,k}^{r} \in \mathbb{C}^{N \times 1}$, and the direct BS-to-user channel vector be $\mathbf{h}_{BU,k}^{r} \in \mathbb{C}^{M \times 1}$. The superscript $r$ denotes the discrete time index corresponding to the $r$-th FL iteration. To align with the subsequent model aggregation stage, we denote the receive beamforming vector at the BS in the $r$-th round as $\mathbf{f}^{r} \in \mathbb{C}^{M \times 1}$, and the RIS phase-shift vector as $\boldsymbol{\theta}^{r} \in \mathbb{C}^{N \times 1}$. The system model is illustrated in Fig~\ref{fig:system}.

To characterize the time-varying channel induced by user mobility, we adopt a decomposition model of large-scale path loss and small-scale fast fading. Let the position of user $k$ in the $r$-th round be $\mathbf{P}_k^{r} \in \mathbb{R}^2$, and let the positions of the RIS and the BS be $\mathbf{P}_R$ and $\mathbf{P}_B$, respectively. The distances from the user to the RIS and the BS are given by $l_{RU,k}^{r} = \norm{\mathbf{P}_k^{r} - \mathbf{P}_R}_2,\ l_{BU,k}^{r} = \norm{\mathbf{P}_k^{r} - \mathbf{P}_B}_2$. The corresponding large-scale path loss coefficients are
\begin{equation}
\begin{aligned}
\beta_{RU,k}^{r}
&= \big( \max(l_{RU,k}^{r}, d_{\min}) \big)^{-2}, \\
\beta_{BU,k}^{r}
&= \big( \max(l_{BU,k}^{r}, d_{\min}) \big)^{-\alpha_d}.
\end{aligned}
\end{equation}
where $d_{\min} > 0$ is the minimum distance threshold and $\alpha_d > 0$ is the path loss exponent of the direct link. The complete channel is then expressed as
\begin{equation}
\mathbf{h}_{x,k}^{r} = \frac{\sqrt{\beta_{x,k}^{r}}}{s_0} \mathbf{g}_{x,k}^{r}, \quad x \in \{RU, BU\},
\end{equation}
where $s_0$ is a normalized reference scale, and $\mathbf{g}_{x,k}^{r}$ denotes the normalized small-scale fading.

The small-scale fading evolves with user mobility. Let $v_k$ be the velocity of user $k$ and $f_c$ the carrier frequency. The Doppler shift is $f_{D,k} = \frac{v_k f_c}{c}$, where $c$ is the speed of light. The temporal correlation is characterized by the Jakes model $\alpha_k(\Delta) = J_0(2\pi f_{D,k} \Delta)$, where $J_0(\cdot)$ is the zero-order Bessel function of the first kind. The small-scale fast fading is then modeled as a first-order AR($1$) process:
\begin{equation}
\begin{aligned}
\mathbf{g}_{x,k}^{r+1}
&= \alpha_k(\Delta T)\mathbf{g}_{x,k}^{r}
 + \sqrt{1-\alpha_k^2(\Delta T)}\,\boldsymbol{\xi}_{x,k}^{r}, \\
&\qquad \boldsymbol{\xi}_{x,k}^{r} \sim \mathcal{CN}(\mathbf{0}, \mathbf{I}).
\end{aligned}
\end{equation}
where $\Delta T$ is the time interval between adjacent FL rounds. This model ensures that the Jakes correlation structure is preserved while providing a discrete-time implementation via the AR(1) recursion.

The pilot observation time generally differs from the uplink aggregation time. We define the uplink time of the $r$-th round as $\tau_r = t_r + \rho \Delta T,\ 0 < \rho < 1$. The small-scale correlation coefficient between the current time $t_r$ and the uplink time $\tau_r$ is $\alpha_{k,\tau} = J_0(2\pi f_{D,k} \rho \Delta T)$. The channel at the uplink time is then generated as
\begin{equation}
\mathbf{g}_{x,k}^{r,\tau} = \alpha_{k,\tau} \mathbf{g}_{x,k}^{r} + \sqrt{1 - \alpha_{k,\tau}^2} \, \boldsymbol{\xi}_{x,k}^{r,\tau},
\end{equation}
where $\boldsymbol{\xi}_{x,k}^{r,\tau} \sim \mathcal{CN}(\mathbf{0}, \mathbf{I})$. After reincorporating the large-scale fading at the corresponding time, we obtain $\mathbf{h}_{RU,k}^{r,\tau}$ and $\mathbf{h}_{BU,k}^{r,\tau}$. This modeling approach provides controllable temporal correlation between the pilot observation at the current time and the subsequent uplink aggregation instant.

Let $\chi_d \in \{0,1\}$ indicate whether the direct link is enabled. Define the effective cascaded channel vector for user $k$ at uplink time in the $r$-th round as $\mathbf{u}_k^{r}\triangleq(\mathbf H_{BR}\mathbf f^{r})^* \odot \mathbf h_{RU,k}^{r,\tau}\in \mathbb{C}^{N \times 1}$. Define $d_k^{r} \triangleq \mathbf f^{rH}\mathbf h_{BU,k}^{r,\tau}$ as the projection of the direct link under the receive beamforming. The corresponding effective scalar channel is 
\begin{equation}\label{eq:eff_c}
c_k^{r}\triangleq(\boldsymbol{\theta}^{r})^{T}\mathbf u_k^{r}+\chi_d\, d_k^{r}.
\end{equation}

During the pilot phase, the RIS applies a phase vector $\boldsymbol{\theta}_p^{r} \in \mathbb{C}^{N}$ in the $p$-th pilot time slot of the $r$-th round. The effective scalar channel in that slot is $c_{k,p}^{r}=(\boldsymbol\theta_{p}^{r})^{T}\mathbf u_k^{r}+ \chi_d\, d_k^{r}$. Let $\mathbf{p}_k^r=[p_{k,1}^r,\dots,p_{k,P}^r]^T \in \mathbb{C}^{P}$ denote the pilot sequence of user $k$ in the $r$-th round, drawn from a DFT-based pilot phase codebook. The received pilot signal at the $p$-th slot is
\begin{equation}
y_{k,p}^{r}=p_{k,p}^{r}c_{k,p}^{r}+n_{k,p}^{r},
\end{equation}
where $n_{k,p}^{r} \sim \mathcal{CN}(0, \sigma_p^2)$ is the pilot noise. Stacking the RIS phase vectors $(\boldsymbol{\theta}_p^{r})^T$ over $p=1,\dots,P$, we define $\boldsymbol{\Theta}^{r}=\big[(\boldsymbol{\theta}_1^{r})^T;\,\cdots;\,(\boldsymbol{\theta}_P^{r})^T\big]\in\mathbb{C}^{P\times N}$. 
The received pilot vector can be written as
\begin{equation}
\mathbf{y}_k^{r}=\operatorname{diag}(\mathbf{p}_k^r)(\boldsymbol{\Theta}^{r}\mathbf u_k^{r}+ \chi_d \, d_k^{r} \mathbf{1}_P)+\mathbf{n}_k^{r},
\end{equation}
where $\mathbf{1}_P$ is an all-one vector of length $P$, and $\mathbf{n}_k^{r} = [n_{k,1}^{r}, \dots, n_{k,P}^{r}]^T$. In this system, the RIS not only improves the quality of the subsequent OTA aggregation link, but also directly determines the generation and statistical structure of the pilot observations.

\subsection{Over-the-Air Model Aggregation}
\label{sec:OTA}
In the $r$-th FL training round (shown in Fig~\ref{fig:system}), all users share the same time-frequency resource to upload their local model updates $\{\Delta \mathbf{w}_k^{r}\}_{k=1}^{K}$, and the BS performs wireless aggregation of their weighted sum via over-the-air computation (AirComp). Let the $d$-th element of the update vector $\Delta \mathbf{w}_k^{r}$ be $\Delta w_k^{r}[d]$, where $1 \leq d \leq D$.

In the AirComp procedure, user $k$ first computes the mean and variance of its local updates: $\mu_k^{r} = \frac{1}{D} \sum_{d=1}^{D} \Delta w_k^{r}[d],\ (\nu_k^{r})^2 = \frac{1}{D} \sum_{d=1}^{D} \big( \Delta w_k^{r}[d] - \mu_k^{r} \big)^2$. The original updates are then normalized to zero-mean, unit-variance normalized symbols $s_k^{r}[d] = \frac{\Delta w_k^{r}[d] - \mu_k^{r}}{\nu_k^{r}},\ 1 \leq d \leq D$. The transmit equalization factor of user $k$ is set to
\begin{equation}
\label{eq:pk}
p_k^{r}=
\omega_k^{r}\sqrt{\eta^{r}}\,\nu_k^{r}\frac{(c_k^{r})^{*}}{|c_k^{r}|^{2}} \in \mathbb{C},
\end{equation}
where $\eta^{r}>0$ is the receiver-side recovery scaling factor (detailed in Section~\ref{sec:Round}). This design ensures that the transmit power satisfies $\mathbb{E} \big[ |x_k^{r}[d]|^2 \big] = |p_k^{r}|^2 \le P_t$, where $P_t$ is the transmission power limit, thereby enabling direct power control while counteracting wireless fading via $p_k^{r}$. The AirComp signal transmitted by user $k$ in the $d$-th time slot is $x_k^{r}[d] = p_k^{r} s_k^{r}[d]$. The signal received at the BS in the $d$-th aggregation time slot is
\begin{equation}\label{eq:ota_receive}
y^{r}[d] = \sum_{k=1}^{K} c_k^{r} p_k^{r}s_k^{r}[d] + n^{r}[d],
\end{equation}
where $n^{r}[d] \sim \mathcal{CN}(0, \sigma_n^2)$ is the additive Gaussian noise. The BS employs a linear reconstructor to estimate the global weighted sum $r^{r}[d] \triangleq \sum_{k=1}^{K} \omega_k^{r} \Delta w_k^{r}[d]$ from $y^{r}[d]$. With inverse compensation of the local mean, the reconstruction is given by
\begin{equation}
\hat{r}^{r}[d] = \frac{1}{\sqrt{\eta^{r}}} y^{r}[d] + \bar{\mu}^{r},
\end{equation}
where $\bar{\mu}^{r} = \sum_{k=1}^{K} \omega_k^{r} \mu_k^{r}$ is the weighted mean compensation term. Stacking the reconstructions over all $D$ time slots yields the aggregated vector estimate  $\hat{\mathbf{r}}^{r} = [\hat{r}^{r}[1], \dots, \hat{r}^{r}[D]]^{T}$.

The AirComp aggregation distortion is measured by the distortion metric $\mathrm{NMSE}_{\mathrm{agg}}^{r}=\norm{\hat{\mathbf{r}}^{r}-\mathbf{r}^{r}}_2^2/\norm{\mathbf{r}^{r}}_2^2$,
which serves as the foundation for the subsequent physical-layer design and system analysis.

\section{Proposed Framework and Algorithm Design}
In this section, we present the key components of our proposed framework. The system is composed of three tightly interacting modules: (i) GRU-based cascaded channel estimation, (ii) Reptile-based personalized federated aggregation, and (iii) dynamics-aware user grouping. These modules are connected through low-dimensional dynamic variables rather than full CSI exchange, ensuring scalability in time-varying RIS-assisted OTA-FL systems. We first introduce the GRU-driven channel estimator in Section~\ref{sec:GRU}, followed by the personalization mechanism in Section~\ref{sec:Personalization}. The per-round FL procedure is then described in Section~\ref{sec:Round}. Finally, the dynamic grouping strategy is developed in Section~\ref{sec:Grouping}, forming a closed-loop design across estimation, communication, and learning.

\subsection{GRU-Based Local Channel Estimation} \label{sec:GRU}
The GRU, introduced in \cite{2014arXiv1406.1078C}, uses update and reset gates to retain low complexity temporal memory, making it suitable for mobility-induced channel dynamics. Under $\mathbf{f}^r$ and $\boldsymbol{\Theta}^r$, the complex pilot observation is transformed as $\mathbf x_k^r = \big[\Re{\mathbf y_k^{r}};\Im{\mathbf y_k^{r}}\big]$ and then fed into the neural estimator.
To improve the physical interpretability of the learning-based estimator, we explicitly embed wireless propagation structure into the model design. Specifically, the channel representation is decomposed into a slowly varying large-scale component and a fast-varying small-scale fading component. This decomposition prevents the entanglement of spatial geometry-induced variations with Doppler-driven temporal fluctuations, which may otherwise degrade temporal consistency in standard autoregressive modeling.

To implement this design, we adopt a shared Conv1D--GRU backbone to extract temporal representations from normalized channel observations, while a user-specific head captures individual large-scale characteristics. Let $\boldsymbol{\phi}^r$ denote the shared parameters, $\boldsymbol{\psi}_k^{r-1}$ the personalized head, and $\mathbf{q}_k^{r-1}$ the recurrent hidden state. The estimator is expressed as
\begin{equation}
\begin{aligned}
(\hat{\mathbf{h}}_{k,t}^r, \Delta\hat{\mathbf{h}}_{k,\tau}^r, \hat{\beta}_k^r)
&= \mathcal{F}(\mathbf{x}_k^r; \boldsymbol{\phi}^r, \boldsymbol{\psi}_k^{r-1}, \mathbf{q}_k^{r-1}), \\
\hat{\mathbf{h}}_{k,\tau}^r
&= \hat{\mathbf{h}}_{k,t}^r + \Delta\hat{\mathbf{h}}_{k,\tau}^r,
\end{aligned}
\end{equation}
where $\hat{\beta}_k^r$ serves as a path-loss proxy for downstream grouping, capturing macroscopic geometric effects.

\begin{figure}[h]
\centering
\includegraphics[width=0.85\linewidth]{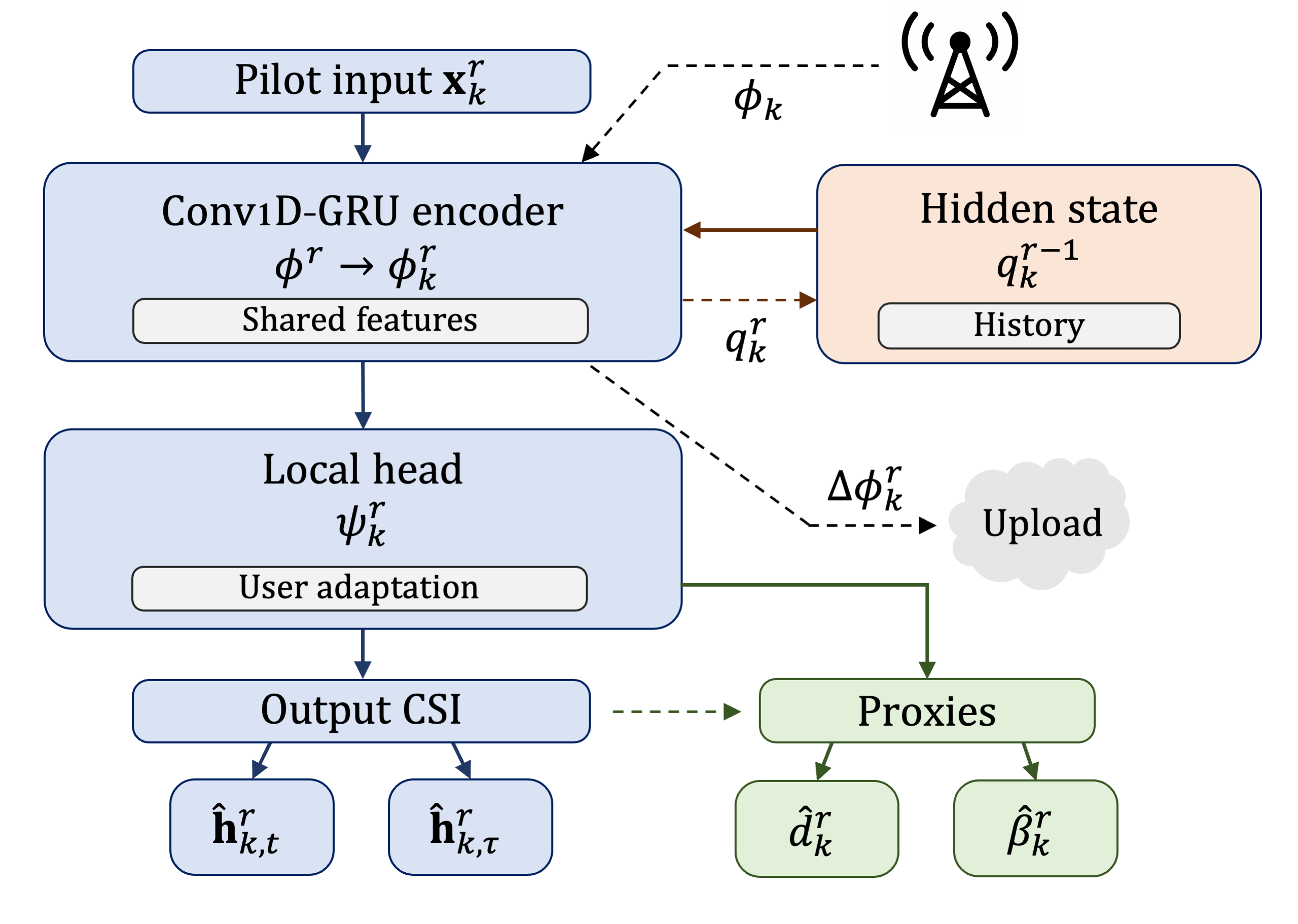}
\caption{Structure of the adopted GRU-based estimator.}
\label{fig:gru}
\end{figure}

To bridge the gap between training supervision and online deployment, the learning procedure is divided into an offline calibration stage and an online inference stage.

\textbf{Offline calibration stage:} The model is trained using reference CSI obtained from linear MMSE (LMMSE)~\cite{10.7551/mitpress/2946.001.0001}. Let $\tilde{\mathbf{h}}_{k,t}^r$ and $\tilde{\mathbf{h}}_{k,\tau}^r$ denote the reference channels. The model is optimized via supervised learning over these calibrated datasets.

\textbf{Online inference stage:} After deployment, reference labels are removed. The trained model operates directly on noisy pilot observations $\mathbf{x}_k^r$ to predict instantaneous CSI, which is then used for over-the-air aggregation and RIS optimization.

The training objective for user $k$ in round $r$ is defined as
\begin{equation}
\begin{aligned}
\mathcal{L}_k^r
&= \frac{1}{2}\big\|\hat{\mathbf{h}}_{k,t}^r - \tilde{\mathbf{h}}_{k,t}^r\big\|_2^2
+ \frac{\lambda_{\alpha,k}^r}{2} \big\|\Delta\hat{\mathbf{h}}_{k,\tau}^r - (\tilde{\mathbf{h}}_{k,\tau}^r - \tilde{\mathbf{h}}_{k,t}^r)\big\|_2^2 \\
&\quad + \lambda_\beta \mathcal{L}_{\beta,k}^r,
\end{aligned}
\end{equation}
where $\mathcal{L}_{\beta,k}^r$ enforces accurate estimation of large-scale path-loss, and $\lambda_{\alpha,k}^r$ and $\lambda_\beta$ control the relative weighting between fast and slow fading components.

We emphasize that the estimator targets the reflection-induced cascaded channel, as RIS chiefly benefits weak or blocked direct links \cite{8647620}, while the stable BS--RIS link makes the user-dependent cascaded part the key estimation burden \cite{9722893}.

\subsection{Shared-Backbone Personalized Federated Updating} \label{sec:Personalization}
We adopt a personalized FL architecture where the model is decomposed into a shared backbone $\boldsymbol{\phi}^r$ and a user-specific head $\boldsymbol{\psi}_k^r$. The backbone captures transferable temporal representations, while the head absorbs user-specific statistical heterogeneity. Each user performs local optimization:
\begin{equation}
(\boldsymbol{\phi}_{k,\mathrm{loc}}^r, \boldsymbol{\psi}_k^r)
= \arg\min_{\boldsymbol{\phi}, \boldsymbol{\psi}_k}
\mathcal{L}_k^r(\boldsymbol{\phi}, \boldsymbol{\psi}_k).
\end{equation}

Only the backbone update is transmitted:
\begin{equation}
\Delta\boldsymbol{\phi}_k^r = \boldsymbol{\phi}_{k,\mathrm{loc}}^r - \boldsymbol{\phi}^r.
\end{equation}

The server adopts a Reptile-style meta-update:
\begin{equation}
\boldsymbol\phi^{r+1}
=\boldsymbol\phi^r+\beta_R\,\hat{\mathbf{r}}^r_{\boldsymbol\phi},
\end{equation}
where $\hat{\mathbf{r}}_{\boldsymbol{\phi}}^r$ is the AirComp-recovered aggregate. The Reptile-based design is adopted due to its lightweight computation and compatibility with communication-constrained OTA aggregation scenarios, where full gradient exchange is infeasible.

\subsection{Physical Layer Optimization and Training Flow} \label{sec:Round}

The uplink aggregation is performed via AirComp using the predicted CSI $\hat{c}_k^r$ obtained from Section~\ref{sec:GRU}. The transmit scaling factor is:
\begin{equation}
p_k^r = \omega_k^r \sqrt{\eta^r} \nu_k^r \frac{(\hat{c}_k^r)^*}{|\hat{c}_k^r|^2},
\end{equation}
where $\hat{c}_k^r$ denotes estimated effective channel. We define:
\begin{equation}
\label{eq:eta_r}
\eta_k^r=\frac{P_t|\hat{c}_k^r|^2}{(\omega_k^r)^2(\nu_k^r)^2}, \qquad 
\eta^r=\min_k \eta_k^r,
\end{equation}
and the corresponding proxy aggregation distortion:
\begin{equation}
\label{eq:proxy_nmse}
\mathrm{NMSE}_{\mathrm{proxy}}^r=
\frac{\sigma_n^2P_t}{\eta^r(\sum_k\omega_k^r)^2}.
\end{equation}

This formulation explicitly reveals the coupling between CSI accuracy and aggregation reliability: improved channel prediction directly enhances $\eta^r$, thereby reducing OTA distortion. The RIS and beamforming are optimized via:
\begin{equation}
\max_{\mathbf f^r,\boldsymbol\theta^r}\ \min_k \eta_k^r 
\quad \text{s.t.}\quad 
\|\mathbf f^r\|_2^2=1,\ \ |\theta_n^r|=1,\ \forall n.
\end{equation}

The problem is solved using successive convex approximation (SCA) or DC-based alternating optimization, both ensuring monotonic improvement. The formulation remains computationally tractable despite non-convexity.

Algorithm~\ref{alg:1round} summarizes round-wise procedure of the training framework under the FL system in Section~\ref{sec:FL}.
\begin{algorithm}[t]
\caption{Per-Round Training Procedure}
\label{alg:1round}
\begin{algorithmic}[1]
\STATE Initialize $\boldsymbol{\phi}^{0}$, $\{\boldsymbol{\psi}_k^{0},\boldsymbol{q}_k^{0}\}_{k=1}^{K}$, $\mathbf f^{0}$, and $\boldsymbol{\theta}^{0}$
\FOR{$r=0,1,\ldots,R-1$}
    \STATE Generate $\boldsymbol{\Theta}_{p}^{r}$ and $\{\mathbf h_{RU,k}^{r},\mathbf h_{RU,k}^{r,\tau},\mathbf h_{BU,k}^{r,\tau}\}_{k=1}^{K}$
    \FOR{each user $k$ in parallel}
        \STATE Construct $\mathbf x_k^{r}$ from pilot observation $\mathbf y_k^{r}$
        \STATE Update $(\boldsymbol{\phi}_{k,\mathrm{loc}}^{r},\boldsymbol{\psi}_k^{r})$ by local training
        \STATE Obtain $\Delta\boldsymbol{\phi}_k^{r}=\boldsymbol{\phi}_{k,\mathrm{loc}}^{r}-\boldsymbol{\phi}^{r}$ and $\hat{\mathbf h}_{RU,k}^{\,r,\tau}$
        \STATE Update hidden state $\boldsymbol{q}_k^{r+1}$
    \ENDFOR
    \STATE Compute $\{\omega_k^{r},\nu_k^{r},\hat{c}_k^{r}\}_{k=1}^{K}$
    \STATE Recover $\hat{\mathbf{r}}^r_{\boldsymbol{\phi}}$ via AirComp
    \STATE Update
    \[
    \boldsymbol{\phi}^{r+1}
    =
    \boldsymbol{\phi}^{r}
    +
    \beta_R \hat{\mathbf{r}}^r_{\boldsymbol{\phi}}
    \]
    \STATE Optimize $(\mathbf f^{r+1},\boldsymbol{\theta}^{r+1})$
\ENDFOR
\STATE \textbf{return} $\boldsymbol{\phi}^{R}$, $\{\boldsymbol{\psi}_k^{R}\}_{k=1}^{K}$, $\mathbf f^{R}$, and $\boldsymbol{\theta}^{R}$
\end{algorithmic}
\end{algorithm}

\subsection{Dynamic Risk Grouping} \label{sec:Grouping}

To mitigate heterogeneity-induced aggregation cancellation, we introduce a dynamics-aware grouping mechanism. The system initially performs global aggregation and switches to grouping once stability is achieved, ensuring fast early convergence and stable late-stage training. We define dynamics proxies:
\begin{equation}
\hat \delta_k^r \triangleq \frac{\|\Delta\hat{ \mathbf h}_{k,\tau}^r\|_2}{\|\hat{\mathbf h}_{k,t}^r\|_2}\ge 0,
\qquad
\hat{\beta}_k^r > 0.
\end{equation}

These form a feature vector:
\begin{equation}
\boldsymbol{\varphi}_k^r= 
\begin{bmatrix} 
-\log \hat\beta_k^r\\ 
\log(1+\hat \delta_k^r) 
\end{bmatrix},
\end{equation}

To incorporate RIS-induced correlation, we define the effective signature of user $k$ as $\mathbf{a}_k^r = (\mathbf{f}_{\mathrm{prev}}^r \mathbf{H}_{BR})^* \odot \hat{\mathbf{h}}_{k,\tau}^r$, where $\mathbf{f}_{\mathrm{prev}}^r$ is the previous-round beamformer. The pairwise compatibility is then
\begin{equation}
\rho_{ij}^r=
\frac{\left|\mathbf a_i^{rH}\mathbf a_j^r\right|}{\|\mathbf a_i^r\|_2\|\mathbf a_j^r\|_2},
\qquad 0\le \rho_{ij}^r\le 1,
\end{equation}
which favors grouping users with correlated RIS signatures and acts as a structural prior for subsequent grouped physical-layer optimization.

We now construct a robust grouping objective. Define a nominal risk score $\bar{r}_k^r = -\log\hat{\beta}_k^r + \lambda_\delta \log(1+\hat{\delta}_k^r)$. To ensure robustness against temporal fluctuations, we further define an uncertainty radius:
\begin{equation}
\begin{aligned}
\epsilon_k^r
&= \gamma_\beta \left|(-\log \hat\beta_k^r)-(-\log \hat\beta_k^{r,\mathrm{EMA}})\right| \\
&\quad + \gamma_\delta \left|\log(1+\hat \delta_k^r)-\log(1+\hat \delta_k^{r,\mathrm{EMA}})\right|,
\end{aligned}
\end{equation}
where the baselines are updated via exponential moving average as $\hat{\beta}_k^{r,\mathrm{EMA}} = \lambda \hat{\beta}_k^{r-1,\mathrm{EMA}} + (1-\lambda)\hat{\beta}_k^r$, and similarly for $\hat{\delta}_k^{r,\mathrm{EMA}}$.

Let $x_k \in [0,1]$ denote the soft membership of user $k$ to the high-risk group $\mathcal{G}_2$, with $1-x_k$ corresponding to the low-risk group $\mathcal{G}_1$. The optimization problem is:
\begin{equation}\label{eq:grouping_p}
\boxed{
\begin{aligned}
\min_{\mathbf x,\mu_1,\mu_2}\quad 
&\sum_{k=1}^K (1-x_k)\Big(|\bar r_k-\mu_1|+\epsilon_k\Big)^2 \\
& + \sum_{k=1}^K x_k\Big(|\bar r_k-\mu_2|+\epsilon_k\Big)^2
-\zeta_1(\mu_2-\mu_1) \\
& + \zeta_2 \sum_{i<j}(1-|x_i-x_j|)(1-\rho_{ij}) \\
& + \zeta_3\sum_{k=1}^K |x_k-x_{k,\mathrm{prev}}| 
+\zeta_4\sum_{k=1}^K x_k(1-x_k) \\
\text{s.t.}\quad 
& 0\le x_k\le 1,\ \forall k, \\
& K_{\min}\le \sum_{k=1}^K x_k\le K-K_{\min}, \\
& \mu_2\ge \mu_1.
\end{aligned}}
\end{equation}

followed by hard assignment $x_k^\star = \mathbb{I}(x_k \ge \tfrac12)$. The coefficients $\zeta_i$ weight group separation, RIS compatibility, temporal stability, and binary regularization, respectively. This problem is solved via successive convex approximation.

To determine when to trigger grouping, we monitor the stability of the dynamic proxies at the population level. Define the population-average dynamic proxy at round $r$ as
\begin{equation}
\bar{\varphi}_i^r = \frac{1}{K} \sum_{k=1}^K \varphi_{k,i}^r, \quad i=1,2,
\end{equation}
where $\varphi_{k,i}^r$ denotes the $i$-th component of $\boldsymbol{\varphi}_k^r$. The corresponding exponentially smoothed aggregate proxy is updated as
\begin{equation}
\varphi_i^{r,\mathrm{EMA}} = \lambda_\varphi \varphi_i^{r-1,\mathrm{EMA}} + (1-\lambda_\varphi) \bar{\varphi}_i^r,
\end{equation}
with $\lambda_\varphi \in (0,1)$ being the smoothing factor. Grouping is activated only when, for both $i \in \{1,2\}$ and thresholds $\varsigma_i$,
\begin{equation}\label{eq:early_stop}
\frac{|\varphi_i^{r,\mathrm{EMA}} - \varphi_i^{r-1,\mathrm{EMA}}|}
{|\varphi_i^{r-1,\mathrm{EMA}}|} \le \varsigma_i
\end{equation}
holds for several consecutive rounds after a prescribed minimum round index. That is, dynamic grouping begins only after the aggregate dynamic proxies have stabilized.

Once activated, each group runs independent OTA-FL and RIS optimization. Beyond reducing aggregation cancellation, the proxies decouple long-range path-loss from short-term dynamics, allowing the shared backbone to focus on scale-free temporal structures. The overall training procedure is summarized in Algorithm~\ref{alg:grouping}.

\begin{algorithm}[t]
\caption{Overall Training Procedure With Dynamic Risk Grouping}
\label{alg:grouping}
\begin{algorithmic}[1]
\REQUIRE $\{\hat{\beta}_k^{r}, \hat{\delta}_k^{r}\}_{k=1}^{K}$
\ENSURE $\{\mathcal G_g^{r}\}_{g=1}^{2}$
\FOR{round $r=0,1,\ldots,R-1$}
    \IF{\eqref{eq:early_stop} is not satisfied}
        \STATE Run 3--13 of Algorithm~\ref{alg:1round} for all users
    \ELSE
        \STATE Solve \eqref{eq:grouping_p} and obtain $\{\mathcal G_g^{r}\}_{g=1}^{2}$
        \FOR{$g=1,2$}
            \STATE Run 3--13 of Algorithm~\ref{alg:1round} for users in $\mathcal G_g^{r}$
        \ENDFOR
    \ENDIF
\ENDFOR
\end{algorithmic}
\end{algorithm}

\section{Performance Analysis}\label{sec:Analysis}
In this section, we analyze the proposed closed-loop framework along three dimensions: (i) convergence behavior of the Reptile-based shared backbone under OTA distortion, (ii) statistical convergence of GRU-driven dynamic proxies, and (iii) convergence of the SCA-based grouping solver. These components jointly characterize the learning--communication--optimization coupling in RIS-assisted OTA-FL systems.

\subsection{Convergence of the Reptile Algorithm for the Shared Backbone}
We analyze the recursion of the shared backbone $\boldsymbol{\phi}^r$ in Section~\ref{sec:Personalization}, where the OTA-recovered update $\hat{\mathbf{r}}_{\boldsymbol{\phi}}^r$ is obtained via AirComp (Section~\ref{sec:Round}).

We decompose the OTA-recovered update as
\begin{equation}\label{eq:ota_e}
\hat{\mathbf{r}}^r_{\boldsymbol{\phi}}=\mathbf{r}_{\boldsymbol{\phi}}^r+\mathbf e_{\mathrm{OTA}}^r,
\end{equation}

where $\mathbf{e}_{\mathrm{OTA}}^r$ captures CSI misalignment and receiver noise. The RIS and beamforming design controls this term through the reliability metric $\eta^r$ and the distortion proxy $\mathrm{NMSE}_{\mathrm{proxy}}^r$ (defined in Section~\ref{sec:Round}).

We impose the following standard assumptions for the first-order meta/FL descent analysis:

\textbf{A1.} $F(\boldsymbol{\phi})$ is $L$-smooth.  

\textbf{A2.} The ideal update $\mathbf{r}_{\boldsymbol{\phi}}^r$ satisfies a descent alignment condition up to bounded heterogeneity bias.  

\textbf{A3.} The OTA perturbation $\mathbf e_{\mathrm{OTA}}^r$ has bounded second moment.

\textit{Lemma 1:}
Let $\mathcal H_r$ denote the filtration up to round $r$.
Under the AirComp model in Section~\ref{sec:Round}, we have
\begin{equation}\label{eq:lemma1_main}
\mathbb E\!\left[\|\mathbf e_{\mathrm{OTA}}^r\|_2^2 \mid \mathcal H_r\right]
\le
C_r\,\mathrm{NMSE}_{\mathrm{proxy}}^r,
\end{equation}
where
\begin{equation}\label{eq:lemma1_Cr}
C_r
=
D_{\boldsymbol{\phi}}
\frac{(\sum_k \omega_k^r)^2}{P_t}
\left(1+\Xi_{\mathrm{CSI}}^r\right),
\end{equation}
and
\begin{equation}\label{eq:Xi_CSI}
\Xi_{\mathrm{CSI}}^r
\triangleq
\frac{P_t}{\sigma_n^2}
\left(
\sum_{k=1}^{K}|\hat c_k^r|^2
\right)
\left(
\sum_{k=1}^{K}|\vartheta_k^r|^2
\right),
\quad
\vartheta_k^r
\triangleq
\frac{c_k^r(\hat c_k^r)^*}{|\hat c_k^r|^2}-1 .
\end{equation}

\textit{Remark:}
$\Xi_{\mathrm{CSI}}^r$ quantifies multiplicative distortion induced by CSI misalignment in AirComp pre-equalization. Under perfect CSI, $\Xi_{\mathrm{CSI}}^r = 0$. Furthermore, if boundedness conditions hold, there exists a constant $\overline C$ such that
\begin{equation}\label{eq:lemma1_uni}
\mathbb E\!\left[\|\mathbf e_{\mathrm{OTA}}^r\|_2^2 \mid \mathcal H_r\right]
\le
\overline C\,\mathrm{NMSE}_{\mathrm{proxy}}^r,
\qquad \forall r.
\end{equation}

\textit{Proof:}
See Appendix~\ref{app:lemma1}.

In particular, increasing $\eta^r$ reduces $\mathrm{NMSE}_{\mathrm{proxy}}^r$ and tightens the OTA perturbation bound. While global convergence remains unproven for general nonlinear neural estimators, they act as bounded-error operators within their intended operating regimes. In this work, the resulting mismatch is absorbed into $\Xi_{\mathrm{CSI}}^r$, which scales only the multiplicative constant $C_r$; hence, if uniformly bounded, it does not alter the descent structure or stationarity guarantee, but merely enlarges the steady-state neighborhood.

Define the meta increment $\Delta\boldsymbol{\phi}^r\triangleq \boldsymbol\phi^{r+1}-\boldsymbol\phi^r$ induced by the Reptile update in Section~\ref{sec:Personalization}. Under $L$-smoothness, the descent lemma gives
\begin{equation}\label{eq:descent_lemma_phi}
F(\boldsymbol\phi^{r+1})
\le
F(\boldsymbol\phi^r)
+
\left\langle \nabla F(\boldsymbol\phi^r),\Delta\boldsymbol{\phi}^r \right\rangle
+
\frac{L}{2}\|\Delta\boldsymbol{\phi}^r\|_2^2.
\end{equation}
Since $\Delta\boldsymbol{\phi}^r$ is proportional to $\hat{\mathbf{r}}^r_{\boldsymbol{\phi}}$, \eqref{eq:ota_e} allows us to interpret $\Delta\boldsymbol{\phi}^r$ as the sum of an ideal descent term and an OTA perturbation term.
Taking conditional expectation given $\boldsymbol\phi^r$, the inner-product term contributes a negative descent component up to a bounded heterogeneity/personalization penalty, while the quadratic term contributes an $O(\beta_R^2)$ disturbance containing both local stochasticity and OTA perturbation energy. Invoking Lemma~1 then yields the following convergence statement.

\textit{Theorem 1:}
Under the above regularity conditions, there exist positive constants $c_i$ determined by the smoothness constant $L$, $\overline C$ and the second-moment growth constant, and the Young's inequality parameter such that
\begin{equation}\label{eq:cond1}
\begin{aligned}
\mathbb E\!\left[F(\boldsymbol\phi^{r+1})\mid \boldsymbol\phi^r\right]
\le\;&
F(\boldsymbol\phi^r)
\,- c_1\|\nabla F(\boldsymbol\phi^r)\|_2^2 \\
&\quad\quad\quad + c_2 \beta_R \underbrace{B_{\mathrm{het}}^r}_{\text{bounded}} \\
&\quad\quad\quad + c_3 \beta_R^2
\underbrace{\sigma_{\mathrm{loc}}^{2,r}}_{\text{bounded}} \\
&\quad\quad\quad + c_4 \beta_R^2\,\mathrm{NMSE}_{\mathrm{proxy}}^r ,
\end{aligned}
\end{equation}
where $B_{\mathrm{het}}^r$ and $\sigma_{\mathrm{loc}}^{2,r}$ denote the personalization bias bound, caused by the Reptile-style meta-updating, and the local stochasticity bound determined by the user-side training data, respectively. Consequently, summing \eqref{eq:cond1} over $r$ implies that $\sum_r \mathbb E\|\nabla F(\boldsymbol\phi^r)\|_2^2$ is bounded up to an additive term proportional to the cumulative disturbance power. For a constant meta step size, the sequence $\{\boldsymbol\phi^r\}$ therefore remain within a stable neighborhood of the stationary set of $F$, with the neighborhood radius scaling with the compounded effect of heterogeneity, local stochasticity, and OTA aggregation distortion. In particular, a positive $c_1$ ensures that gradient descent drives the learning process. The shared backbone improves FL aggregation consistency and, together with risk grouping, mitigates $\beta_R B_{\mathrm{het}}^r$. Meanwhile, improving $\eta^r$ and reducing $\mathrm{NMSE}_{\mathrm{proxy}}^r$ tighten the steady-state neighborhood.

\textit{Proof:}
See Appendix~\ref{app:theorem1}.

Once risk grouping is activated (Section~\ref{sec:Grouping}), each group runs the OTA-FL procedure on a more homogeneous user subset, which reduces cross-user update cancellation and effectively decreases the heterogeneity-related disturbance terms in Theorem~1. This explains the empirically observed ``fast warmup then stable refinement'' behavior described in Section~\ref{sec:Grouping}.

\subsection{Statistical Convergence of the GRU-Driven Dynamic Proxies}
Due to the nonlinear nature of GRU-based estimators, direct analytical convergence is intractable. Instead, we adopt a statistical convergence perspective by tracking dynamic proxy stability. We analyze three indicators: (i) path-loss proxy stabilization, (ii) mobility proxy consistency, and (iii) cross-user separability. 

\begin{figure*}[!t]
\centering
\subfloat[]{\includegraphics[width=0.36\textwidth]{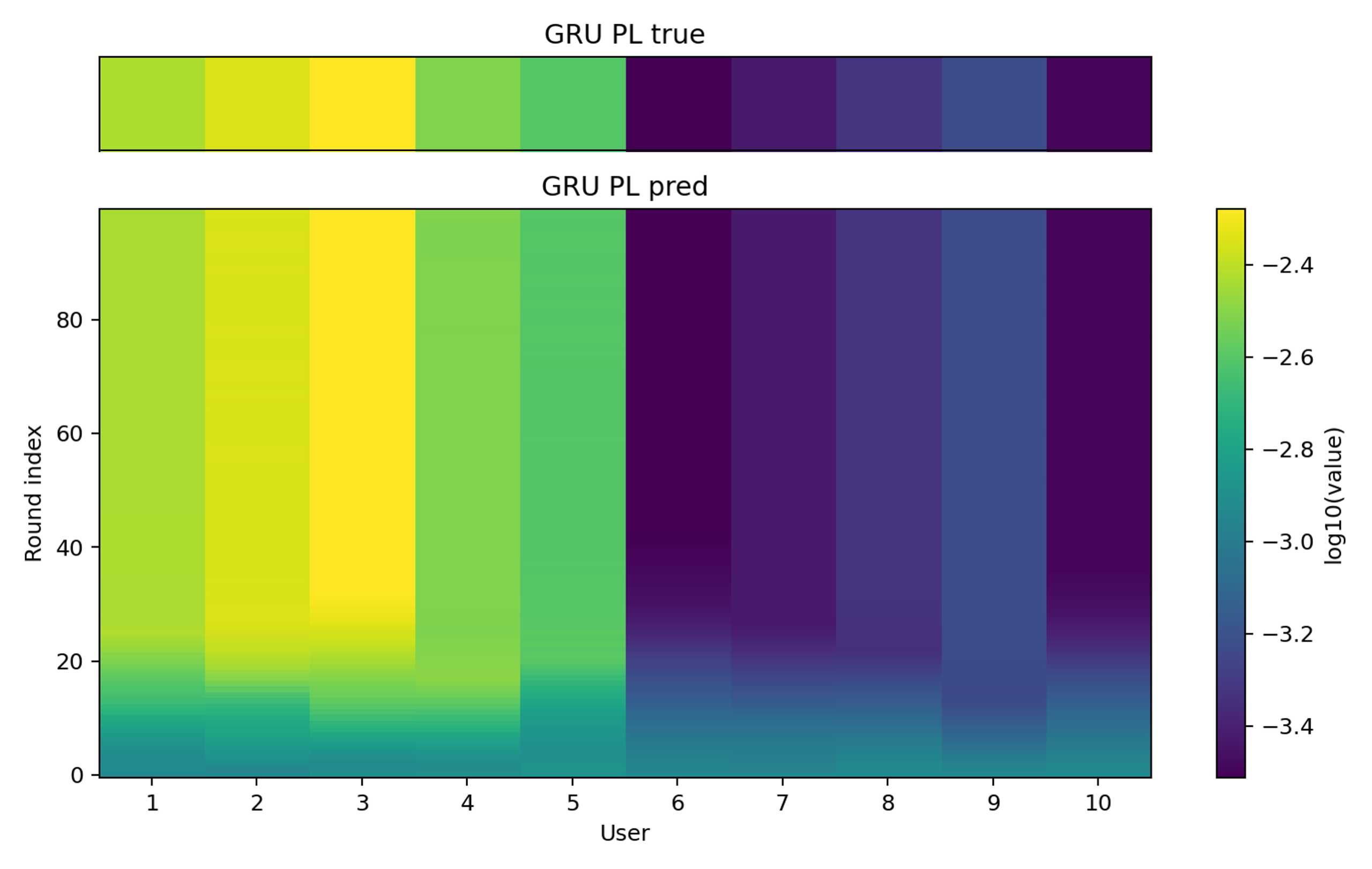}}
\hfill
\subfloat[]{\includegraphics[width=0.36\textwidth]{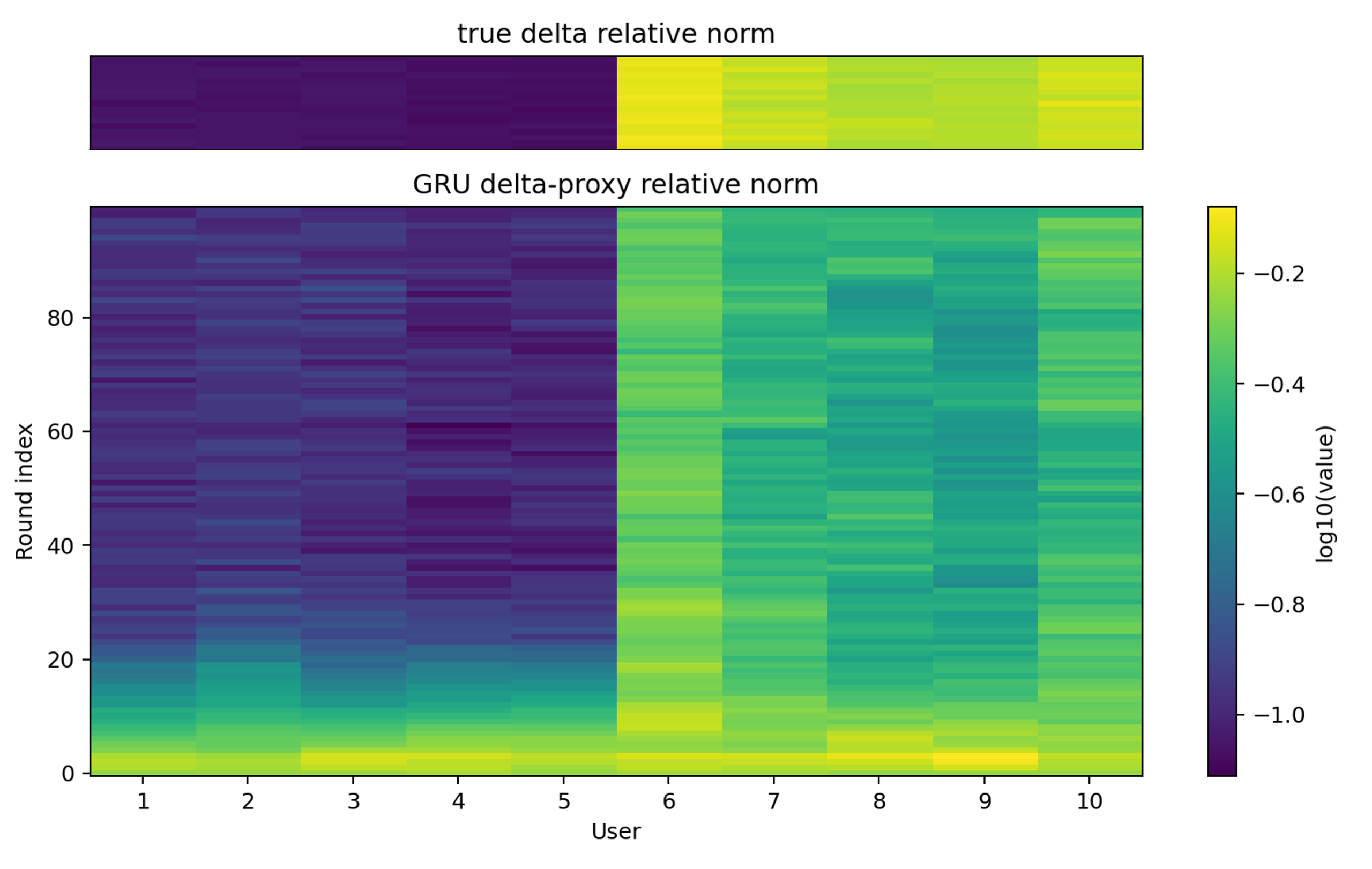}}
\hfill
\subfloat[]{\includegraphics[width=0.27\textwidth]{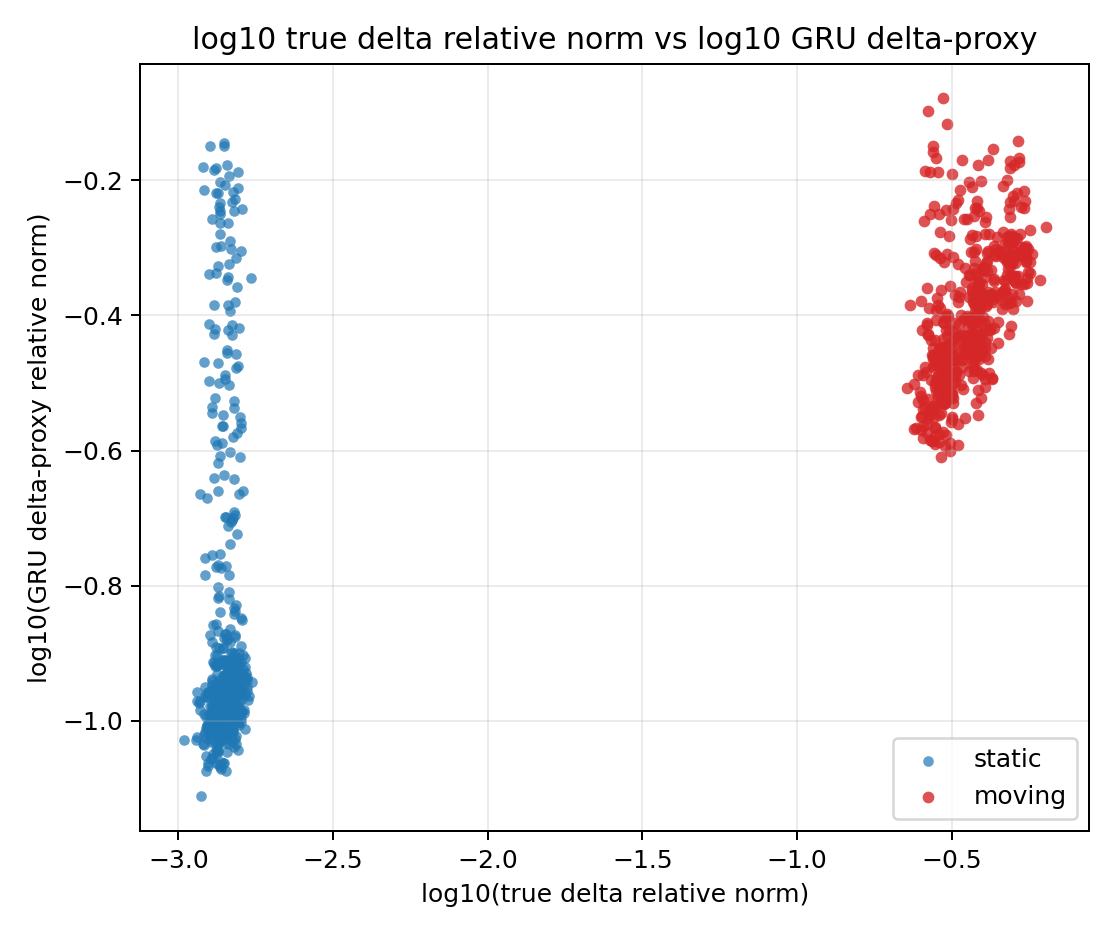}}
\caption{Dynamic/physical proxies across users. (a) Path-loss proxy heatmap. (b) Short-range dynamics proxy norm heatmap. (c) Converged short-range dynamics proxy scatter.}
\label{fig:proxy_maps}
\end{figure*}

As shown in Fig.~\ref{fig:proxy_maps}, all proxies converge after approximately $20$ rounds, indicating stable representation learning. In particular, mobile and static users become clearly separable in the learned embedding space. This stabilization directly enables reliable activation of the grouping module in Section~\ref{sec:Grouping}. These diagnostics complement theoretical convergence by empirically verifying representation stability under time-varying wireless conditions.

\subsection{Convergence of the SCA Grouping Solver}
Let $J(\mathbf x,\mu_1,\mu_2)$ denote the relaxed grouping objective in Section~\ref{sec:Grouping}, and define $z\triangleq(\mathbf x,\mu_1,\mu_2)$. The SCA grouping solver generates a sequence $\{z^{(m)}\}$ by solving, at each iteration, a convex surrogate $\hat J^{(m)}(z)$ constructed at the current point $z^{(m)}$. The nonconvexity of $J$ arises from the bilinear group-fitting terms, the RIS-compatibility coupling involving $|x_i-x_j|$, and the binary-pushing regularizer $x_k(1-x_k)$. Following the standard majorization and first-order consistency principles of SCA/BSUM methods \cite{doi:10.1137/120891009}, the surrogate is chosen as a valid upper approximation of $J$ that is tight at the current iterate and gradient-consistent with the smooth components of the original objective. Under standard regularity conditions, including compact feasibility and a suitable constraint qualification, the resulting objective sequence $\{J(z^{(m)})\}$ is monotonically nonincreasing and convergent. Moreover, every accumulation point of $\{z^{(m)}\}$ is a stationary (KKT) point of the original constrained grouping problem \cite{7776948}.

Since the relaxed grouping problem remains nonconvex, the above guarantee should be interpreted as convergence to a local KKT-stationary solution rather than a globally optimal partition. This suffices for the proposed system design: the hysteresis term and the binary-pushing regularizer are introduced not to convexify the problem, but to make the obtained stationary grouping solution temporally stable and robust against oscillatory regrouping, thereby supporting reliable late-stage grouped aggregation.

\section{Numerical results}\label{sec:Results}
\subsection{Simulation Settings}\label{sec:setting}
We construct the simulation code based on the foregoing content and settings below\footnote{Available at \url{https://github.com/Riko-Neko/gru-meta-aircomp-fl}.}. We consider a RIS-assisted uplink system where the BS is equipped with $M=32$ antennas, the RIS consists of $N=64$ reflecting elements, and $K=10$ users fully participate in each communication round. For the GRU-based estimator, each training task is constructed over a sliding time window of length $W=8$, and the default pilot length is $P=8$. The time-varying channel follows the AR(1) Jakes-correlated model in Section~\ref{sec:Pilot} with carrier frequency $f_c=3.5$~GHz and inter-round sampling interval $\Delta T=10^{-3}$~s. The uplink instant within each round is set to $\rho=0.5$, and the pilot SNR is fixed to $30$~dB. Reference supervision labels are generated using pilot symbols $P_{\mathrm{ref}}=64$ by LMMSE. Unless otherwise specified, reference labels are used instead of perfect labels, and only the reflected link is enabled.

For federated training, we run up to $R=100$ communication rounds. Each user performs $E=3$ local epochs per round with batch size $B=8$, local learning rate  $\beta_{\mathrm{loc}}=10^{-3}$, and Reptile outer step size $\beta_R=0.2$. For OTA aggregation, the uplink SNR is set to $0$~dB and the transmit power is $P_t=0.1$. On the server side, the receive beamformer and RIS phase shifts are updated using the SCA-based solvers with at most $20$ iterations per round.

We construct two scene layouts and three mobility settings to model heterogeneous spatial-temporal conditions. The BS and RIS are located at $(0,0)$ and $(30,0)$~m, respectively. 

Users are distributed around reference points with random offsets up to $\pm 10$~m. 

\begin{figure}[t]
\centering
\includegraphics[width=0.9\linewidth]{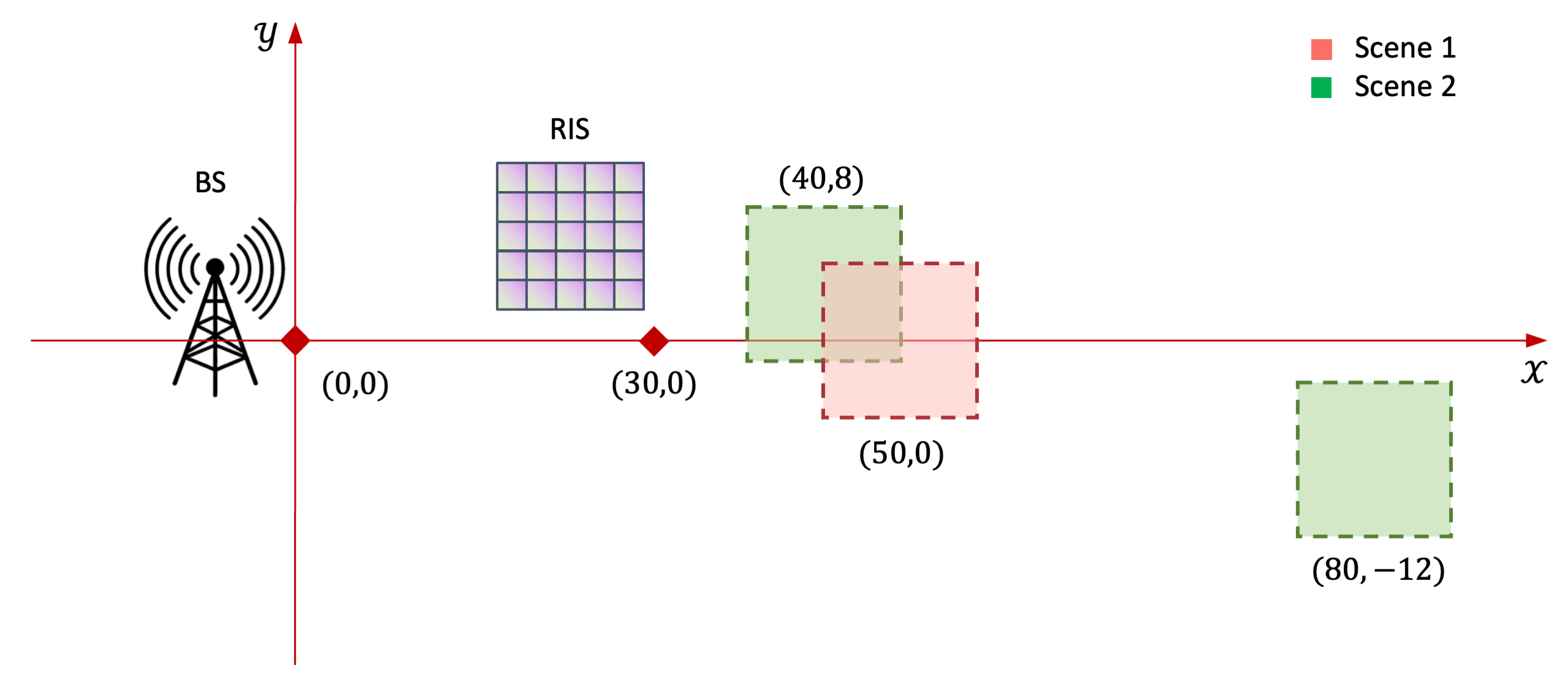}
\caption{Scene-distribution designs.}
\label{fig:scene_design}
\end{figure}

The scene layout, illustrated in Fig.~\ref{fig:scene_design}, controls large-scale heterogeneity:
\begin{itemize}
    \item \textbf{Scene 1 (homogeneous user distribution):} 
    All users are located around $(50,0)$~m, resulting in relatively small path-loss variation.

    \item \textbf{Scene 2 (near/far heterogeneity):} Users are divided into near and far groups located around $(40,8)$~m and $(80,-12)$~m, respectively, producing significant large-scale channel disparity.
\end{itemize}

The mobility settings control temporal dynamics:
\begin{itemize}
    \item \textbf{Setting 1 (mild mobility):} Half of users move at $0.5$--$1.5$~m/s; others are static.

    \item \textbf{Setting 2 (tail-user mobility):} Two far users move at $2.5$--$5$~m/s; others are static.

    \item \textbf{Setting 3 (high mobility):} Random users move at $5$--$10$~m/s, and tail users move at $10$--$20$~m/s.
\end{itemize}

We evaluate the uplink channel prediction accuracy using the round-wise uplink $\mathrm{NMSE}_{\mathrm{up}}^r=\frac{\sum_{k=1}^K \norm{\hat{\mathbf h}_{RU,k}^{\,r,\tau}-\mathbf h_{RU,k}^{r,\tau}}_2^2}{\sum_{k=1}^K \norm{\mathbf h_{RU,k}^{r,\tau}}_2^2}$, and aggregation quality using $\mathrm{NMSE}_{\mathrm{agg}}^r$ and proxy distortion $\mathrm{NMSE}_{\mathrm{proxy}}^r$. We apply a common-reference normalization for the proxy distortion to ensure fair comparison. 

\subsection{Compared Methods}
We compare the following baselines:
\begin{itemize}
    \item \textbf{GRU (proposed):}  Conv1D + GRU backbone (hidden size 32) with personalized head for CSI prediction.
    
    \item \textbf{CNN-arch:} Replaces the GRU backbone with a non-recurrent CNN backbone while keeping the same FL/OTA pipeline, using Conv1D, FC (32), and last-step temporal pooling.
    
    \item \textbf{CNN-base:}  Pure CNN-based federated CSI estimator following \cite{9625822}, using latest-step input with Conv1D and FC layers.
    
    \item \textbf{LMMSE:} Classical LMMSE estimator.
\end{itemize}

\subsection{Main Results}\label{sec:main_results}
\begin{figure}[t]
\centering
\includegraphics[width=0.9\linewidth]{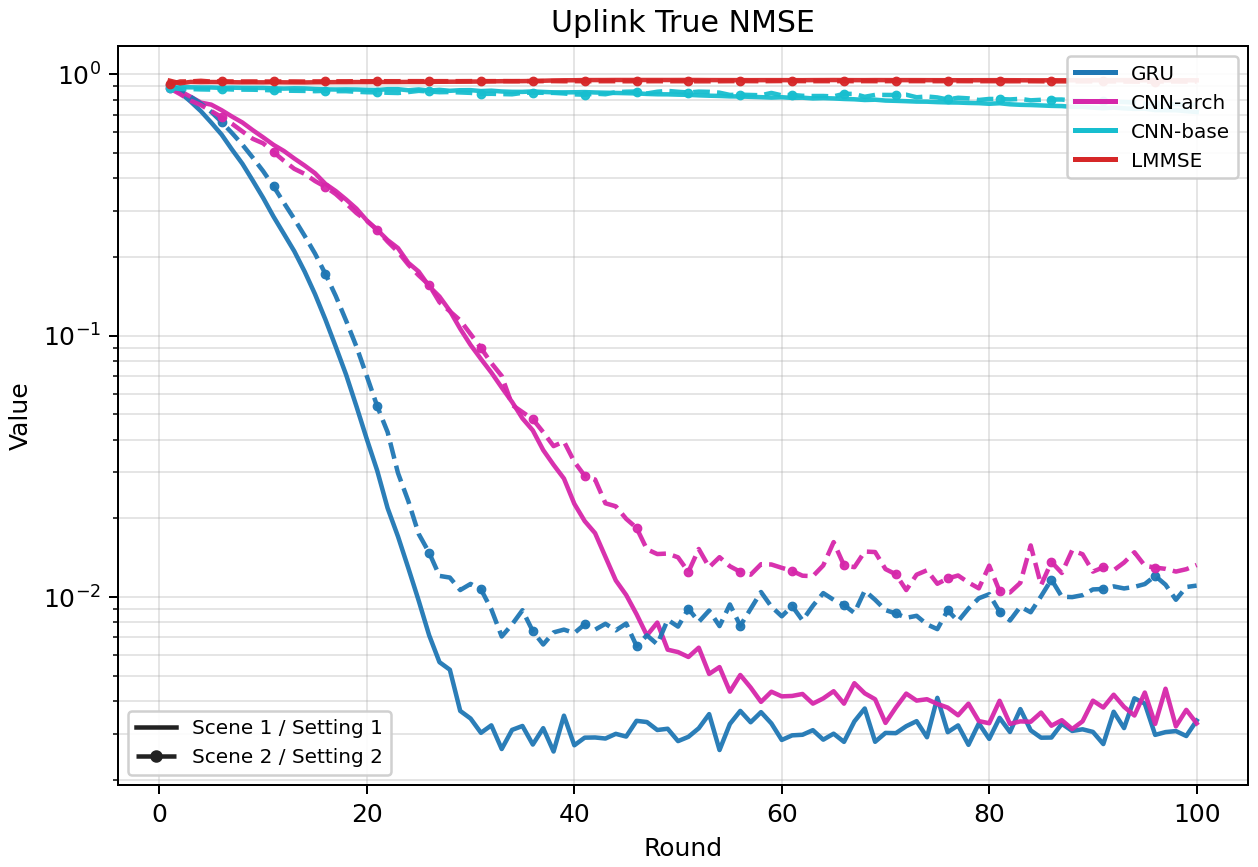}
\caption{Round-wise uplink true NMSE in Scene~1 with Setting~1 and Scene~2 with Setting~2 (without grouping).}
\label{fig:uplink_s1set1_s2set2}
\end{figure}

We first evaluate uplink prediction performance under Scene~1 with Setting~1 and Scene~2 with Setting~2. As shown in Fig.~\ref{fig:uplink_s1set1_s2set2}, GRU consistently achieves the lowest NMSE across both scenarios. It is observed that under near/far heterogeneity and tail mobility, the convergence NMSE is higher than in the mild scenario. This indicates that temporal dynamics and user heterogeneity jointly increase the difficulty of CSI prediction in non-stationary environments.

Table~\ref{tab:best_lr} reports the steady-state uplink NMSE averaged over rounds 50--100 under individually tuned learning rates in Scene~1 with Setting~1. GRU achieves the lowest average NMSE, confirming that its gain is not attributable solely to learning-rate selection.

\begin{table}[t]
\caption{Steady-state NMSE under the best learning rate}
\label{tab:best_lr}
\centering
\begin{tabular}{lcc}
\toprule
Method & Best learning rate & Mean NMSE (Rounds 50--100) \\
\midrule
GRU      & $\beta_{\mathrm{loc}} = 1.05 \times 10^{-3}$ & $3.26 \times 10^{-3}$ \\
CNN-arch & $\beta_{\mathrm{loc}} = 1.05 \times 10^{-3}$ & $3.66 \times 10^{-3}$ \\
CNN-base & $\beta_{\mathrm{loc}} = 9.60 \times 10^{-3}$ & $6.15 \times 10^{-2}$ \\
LMMSE    & -- & $9.50 \times 10^{-1}$ \\
\bottomrule
\end{tabular}
\end{table}

\begin{figure}[t]
\centering
\includegraphics[width=0.9\linewidth]{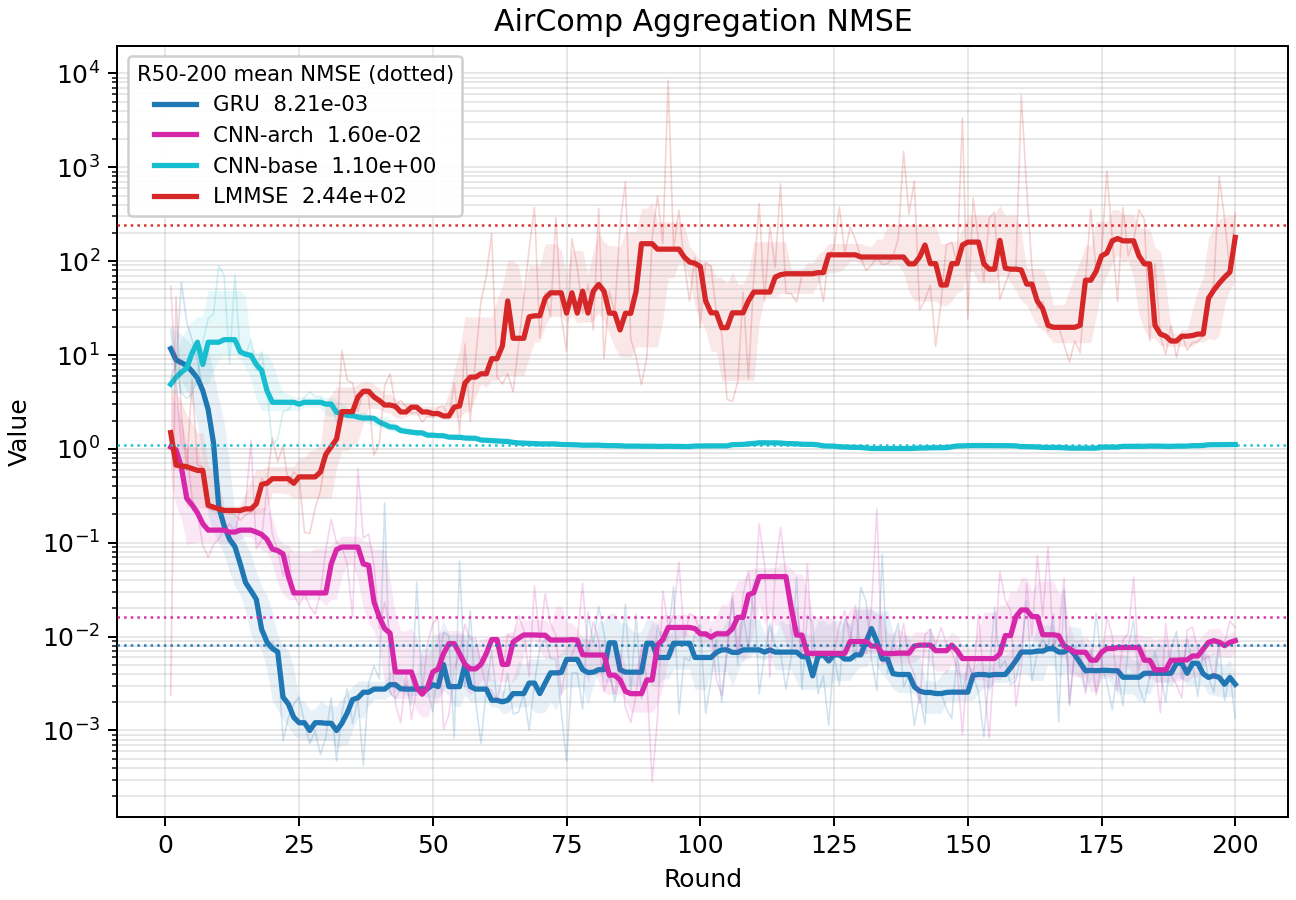}
\caption{Round-wise AirComp aggregation NMSE in Scene~2 with Setting~1 (without grouping).}
\label{fig:aircomp_nmse_s2set1}
\end{figure}

\begin{figure}[t]
\centering
\includegraphics[width=0.9\linewidth]{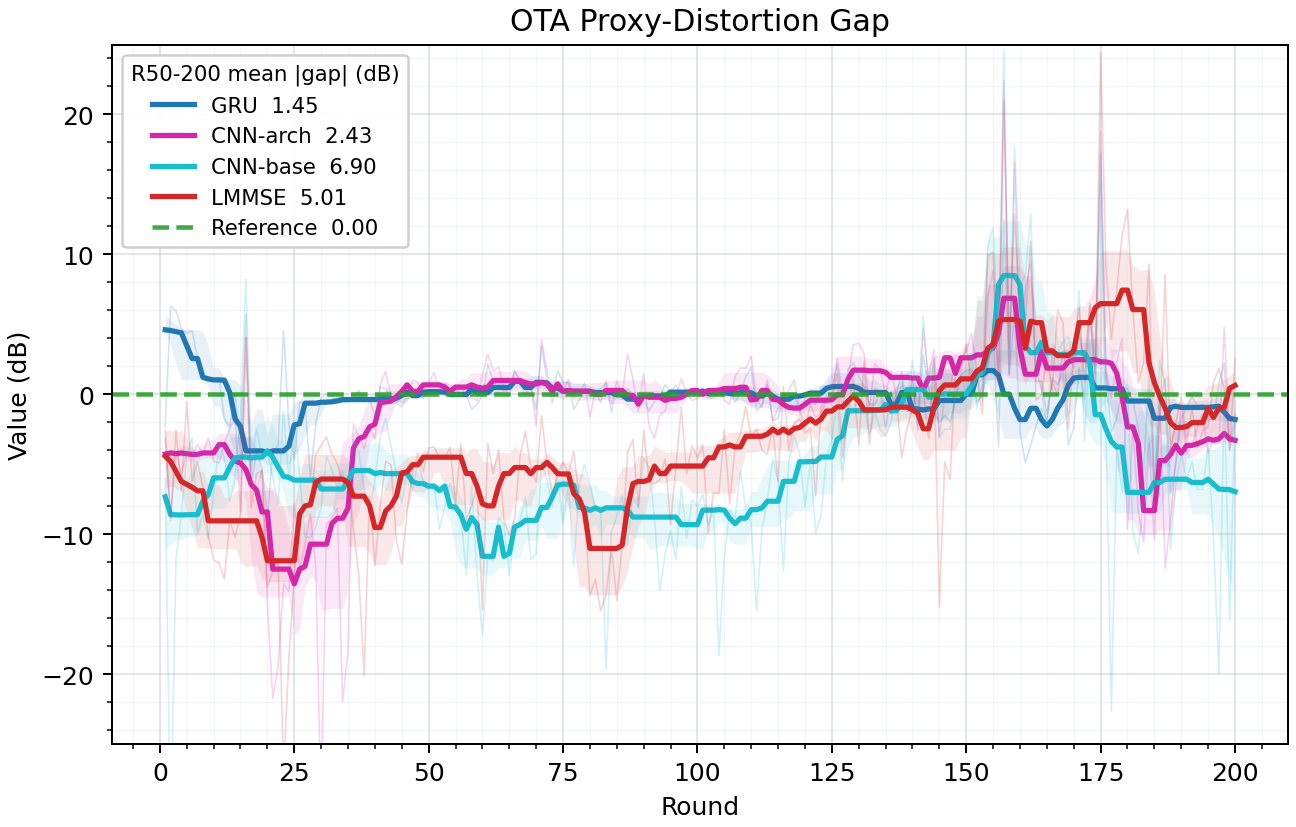}
\caption{Round-wise proxy-distortion gap after physical-layer optimization in Scene~2 with Setting~1 (without grouping).}
\label{fig:proxy_nmse_s2set1}
\end{figure}

We next evaluate AirComp aggregation performance in Scene~2 with Setting~1 over 200 communication rounds. As shown in Fig.~\ref{fig:aircomp_nmse_s2set1}\footnote{Both Fig~\ref{fig:aircomp_nmse_s2set1} and Fig~\ref{fig:proxy_nmse_s2set1} overlay 11-round rolling medians, with mean values computed from round 50 onward.}, GRU achieves consistently lower aggregation NMSE compared to all baselines in the steady state. The early-stage improvement is attributed to more aligned model updates, while later fluctuations arise from reduced update magnitude and increased heterogeneity effects.

Fig.~\ref{fig:proxy_nmse_s2set1} reports the proxy-distortion gap relative to the reference-label configuration under common-reference normalization. Values near zero indicate close agreement with reference-label optimization. GRU reaches a smaller steady-state gap faster, indicating better alignment of estimated-CSI-driven optimization with the reference.

\begin{figure}[t]
\centering
\includegraphics[width=0.9\linewidth]{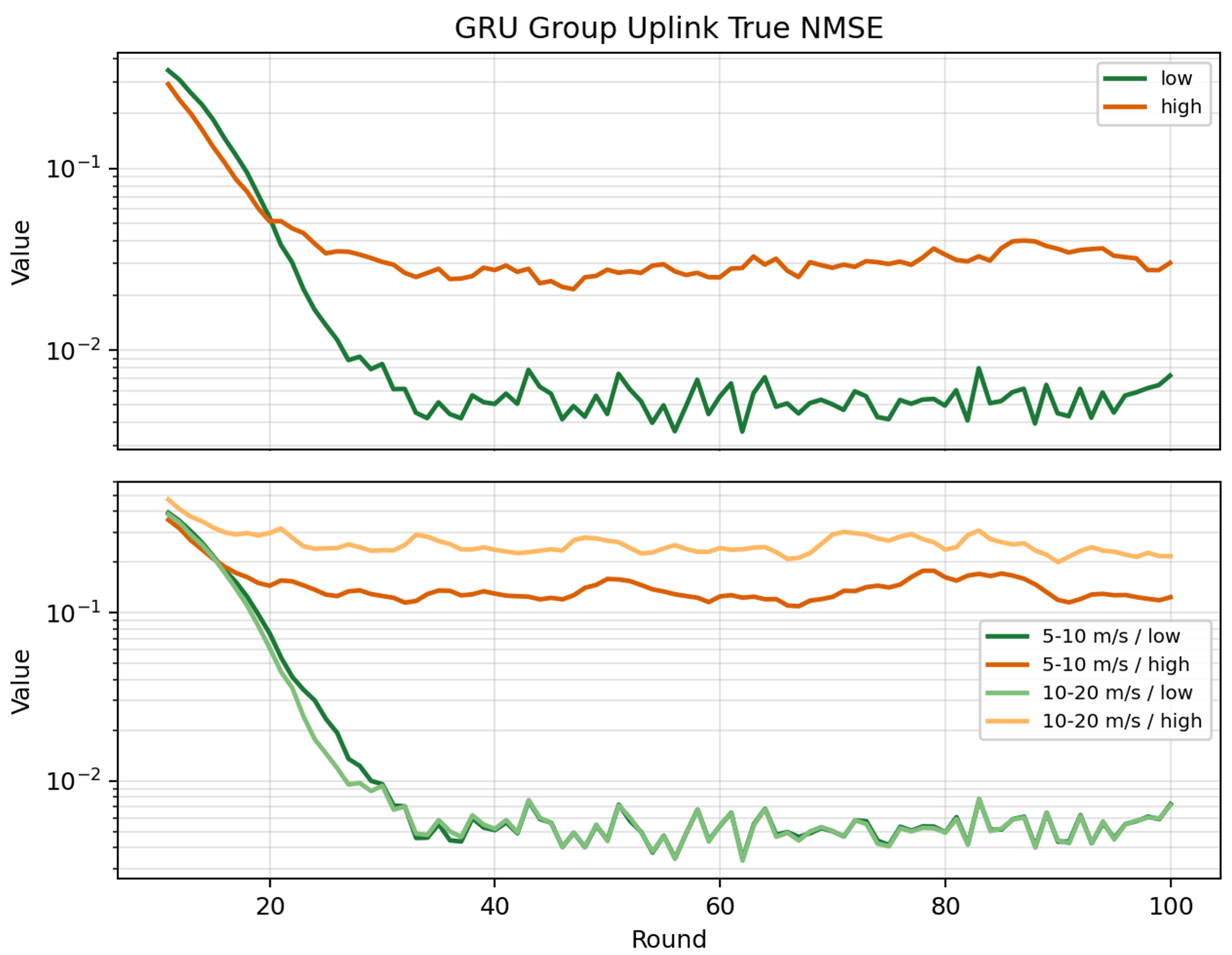}
\caption{Effect of grouping on GRU uplink true NMSE in Scene~2 with Setting~2 (top) and Setting~3 (bottom).}
\label{fig:group_uplink_s2set2_s2set3}
\end{figure}

We further evaluate grouping under Scene~2 with high mobility settings. As shown in Fig.~\ref{fig:group_uplink_s2set2_s2set3}, grouping improves performance by separating users with different channel dynamics. In particular, low-risk users achieve lower NMSE levels, while high-risk users remain at higher error floors due to more challenging channel conditions.

Overall, the proposed framework shows consistent improvements under both mild and highly heterogeneous scenarios, indicating robustness of the joint learning, communication, and optimization design.
\subsection{Further Discussion}\label{sec:further}
\begin{figure}[t]
\centering
\includegraphics[width=0.9\linewidth]{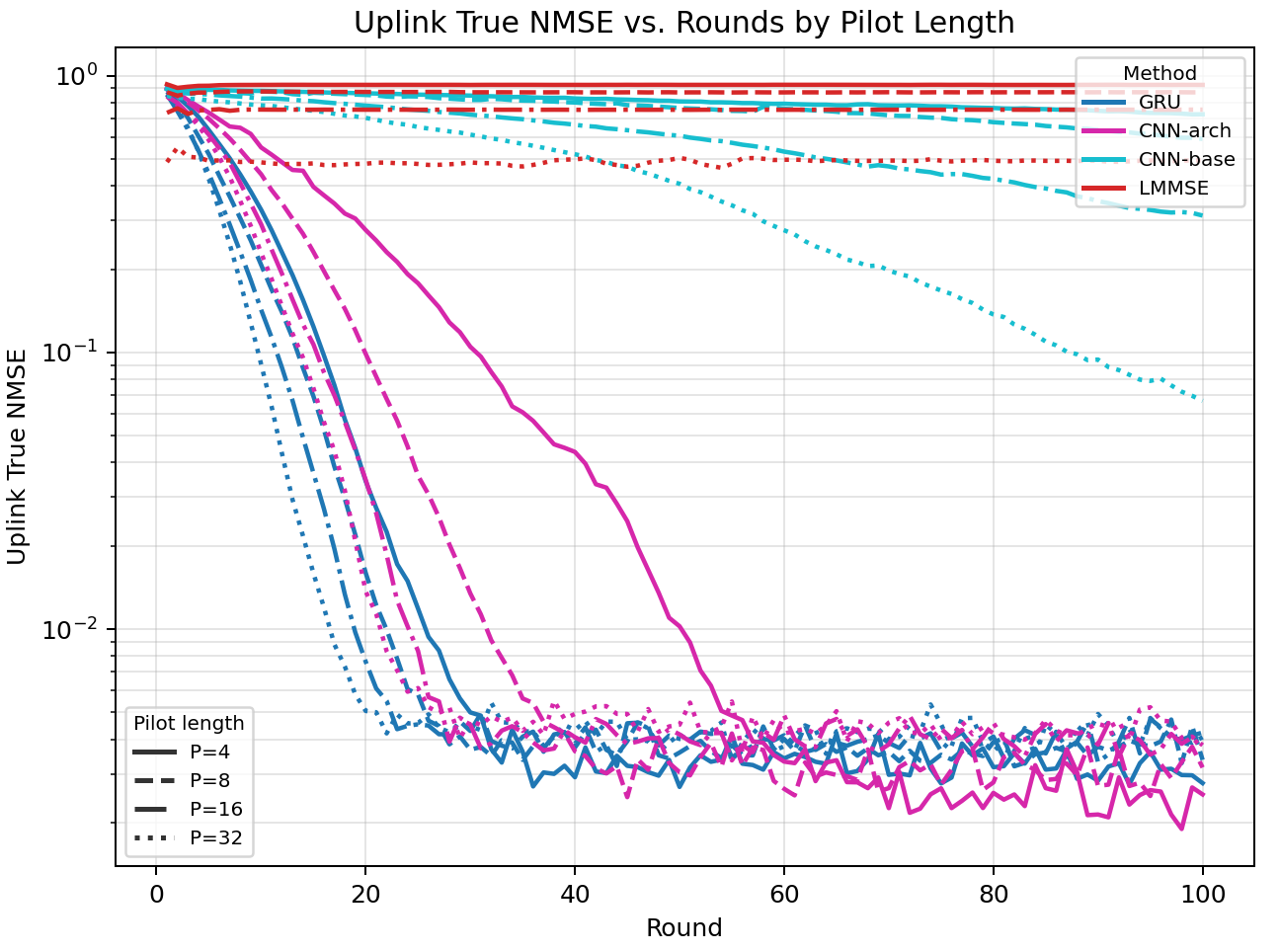}
\caption{Final uplink true NMSE versus pilot length in Scene~1 with all users static.}
\label{fig:pilot_length}
\end{figure}

Fig.~\ref{fig:pilot_length} evaluates performance under different pilot lengths in Scene~1. The results show that GRU achieves larger gain gaps between other methods in low-pilot regimes, indicating stronger robustness when CSI observations are limited, due to the historical information exploitation capability from temporal-memory design.

\begin{figure}[t]
\centering
\includegraphics[width=0.9\linewidth]{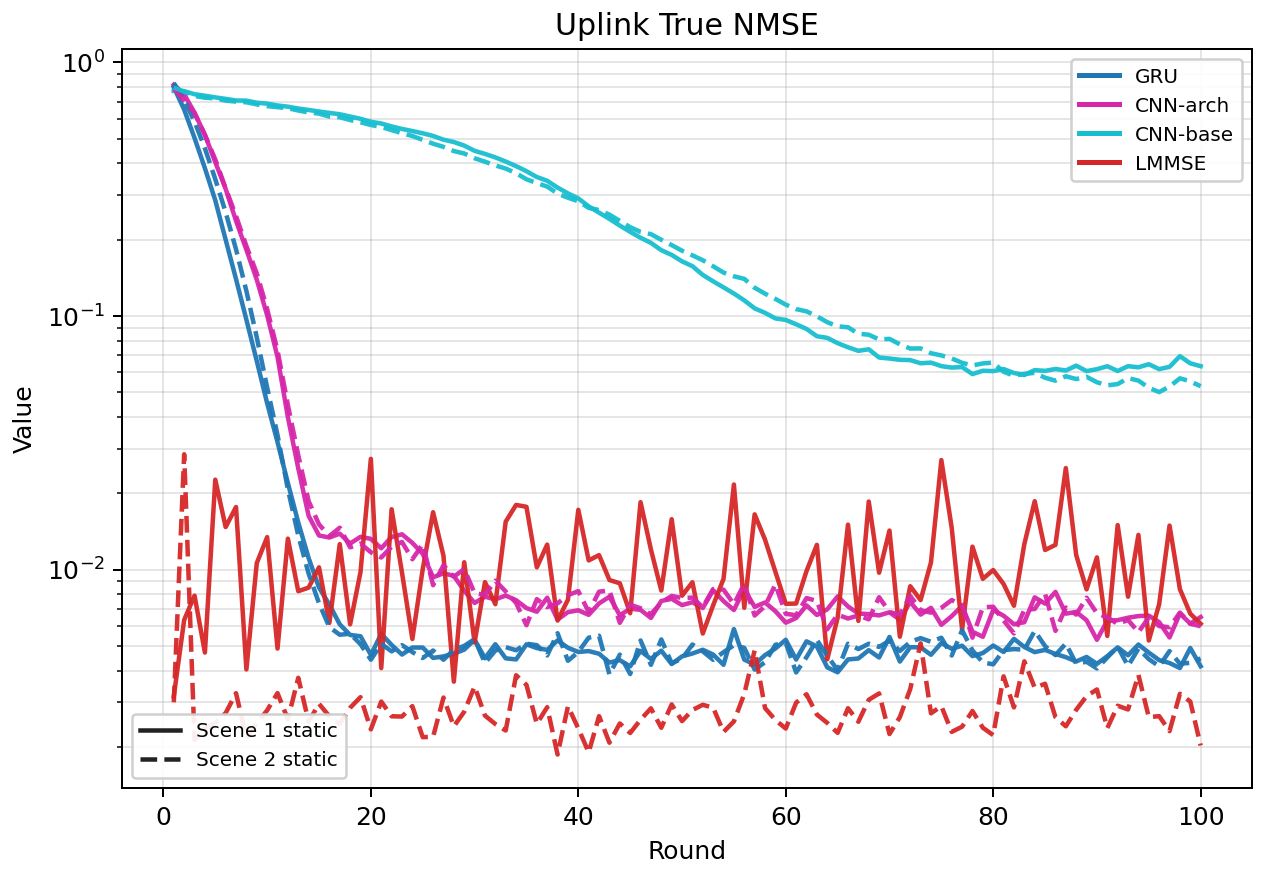}
\caption{Round-wise uplink true NMSE in the extreme pilot case $P=128$ for Scene~1 and Scene~2 with all user static.}
\label{fig:p128}
\end{figure}

We show a large pilot case with $P=128$ as shown in Fig.~\ref{fig:p128}, LMMSE achieves competitive CSI estimation performance; however, its advantage does not fully translate into improved AirComp performance, indicating that CSI accuracy alone is insufficient for optimal system-level optimization.

\begin{figure}[t]
\centering
\includegraphics[width=0.9\linewidth]{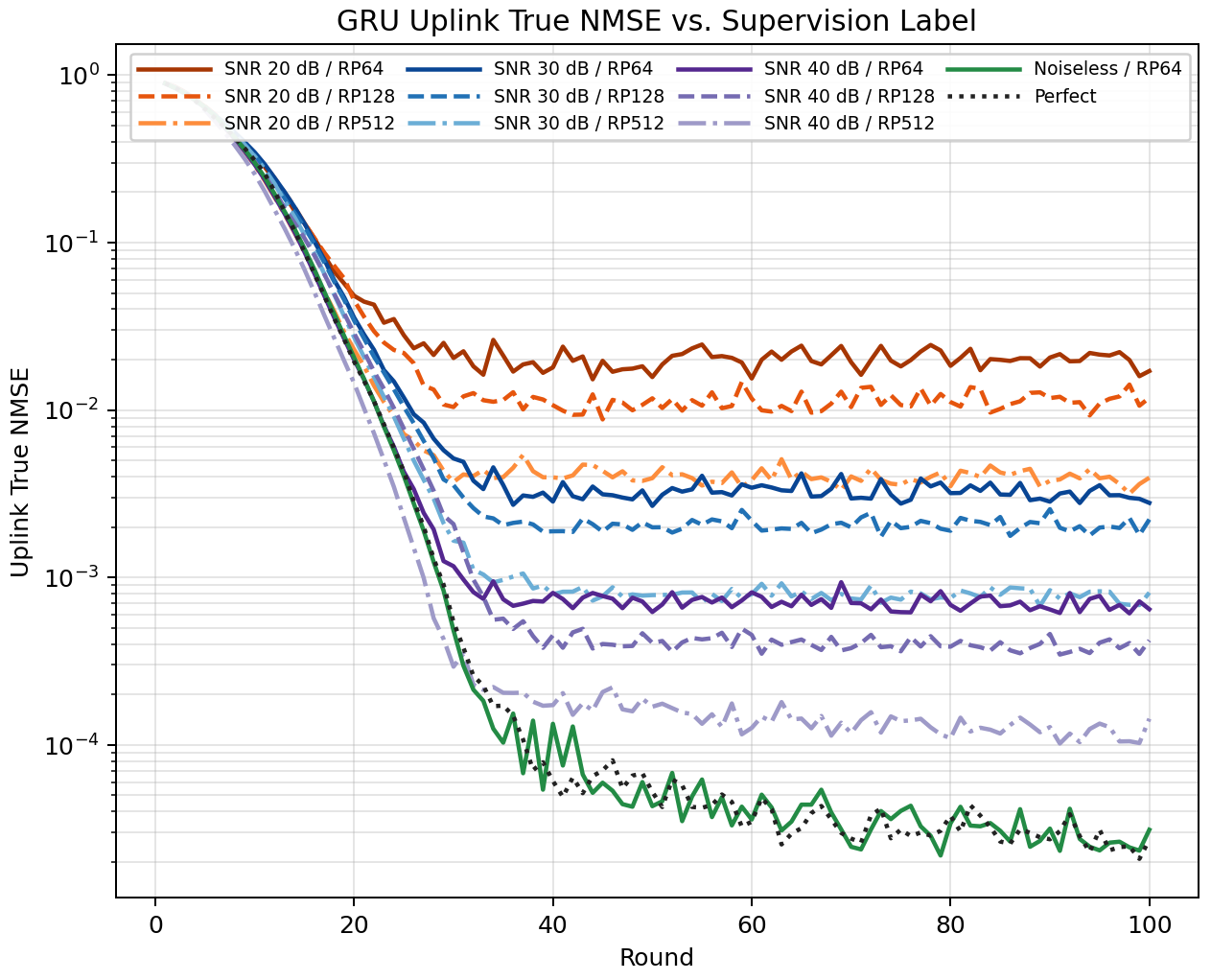}
\caption{GRU uplink true NMSE under different supervision-label qualities in Scene~1 with all users static.}
\label{fig:label_source}
\end{figure}

Fig.~\ref{fig:label_source} evaluates the impact of supervision-label quality on GRU channel estimation in Scene~1 with all users static. Increasing either the reference-pilot SNR or the reference-pilot length consistently reduces the uplink true NMSE. Moreover, the noiseless reference estimate with 64 pilots approaches the performance obtained using perfect CSI labels, demonstrating that label-estimation error is the primary source of the remaining performance gap.

Angle-domain channel estimation (ADCE) in \cite{9625822} generates supervision labels via sparse reconstruction from high-dimensional array observations. In contrast, our labels are inferred from beam-combined pilots, avoiding angular dictionaries and iterative support recovery to reduce pilot, memory, and computational overhead, at the cost of unrecoverable spatial information and greater sensitivity to noise and sensing-matrix conditioning.

\begin{figure}[t]
\centering
\includegraphics[width=0.9\linewidth]{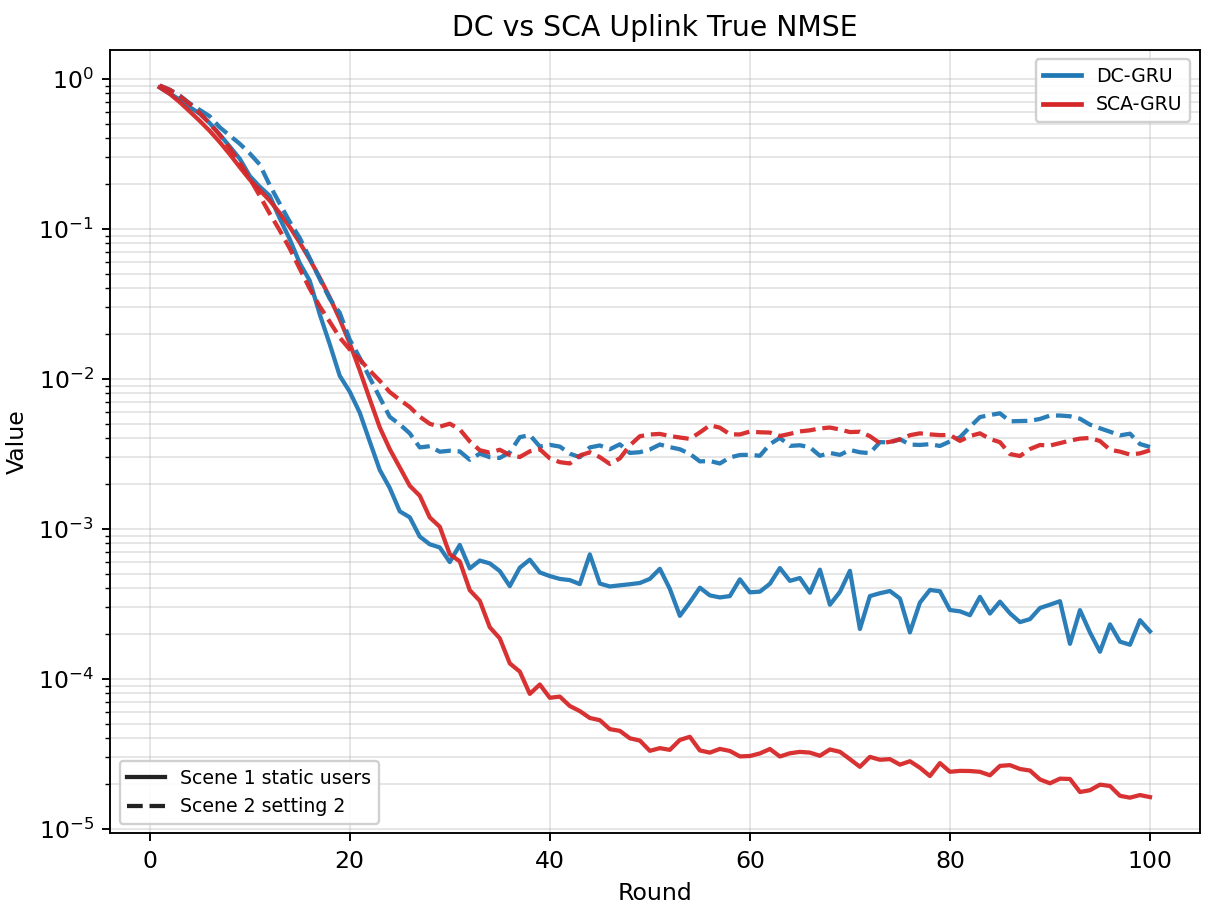}
\caption{DC-AO versus SCA in Scene~1 with all users static and Scene~2 with Setting~2.} 
\label{fig:solver_s1s2stas2}
\end{figure}

We also compare the SCA and DC-AO solvers in Fig.~\ref{fig:solver_s1s2stas2} under perfect supervision label to explicitly illustrate how CSI quality could affect physical layer optimization. SCA performs better in homogeneous settings, while both methods exhibit similar performance under severe heterogeneity, where system performance is primarily limited by channel conditions rather than optimization accuracy.

We further clarify the distinction between dynamic grouping and conventional device selection. When an extremely poor user appears, device selection would typically drop it in the current round, whereas risk grouping keeps it active in the high-risk group. Hence, even when the high-risk group cannot reach a low error floor, its partial convergence remains meaningful because these difficult users are retained rather than discarded. The key difference is semantic rather than functional: grouping preserves full participation and continued support for difficult users, and allows re-admission once their path-loss condition or short-term dynamics improve, whereas device selection achieves robustness by excluding them.

From a mathematical perspective, the benefit of grouping follows from feasible-set inclusion: the grouped problem contains single global aggregation as a special case, and hence its optimum cannot be worse. The gain comes from replacing one compromise model for heterogeneous users with group-wise shared models that better match intra-group statistics, thereby reducing heterogeneity-induced fitting bias.

We additionally note that the direct link shows negligible impact under the considered simulation settings, as its contribution can be largely absorbed by beamforming and RIS-assisted reflection optimization.

\subsection{Complexity Discussion}
The AirComp uplink complexity scales linearly with update dimension $D$, making it communication-efficient for large-scale FL systems. SCA-based joint optimization incurs complexity $O(I_{\max}K^3)$, while the DC-AO alternative has complexity $O(I_{\max}(M^6+N^6))$ due to iterative matrix lifting and difference-of-convex optimization.

The grouping SCA subproblem involves $O(K^2)$ pairwise terms. A worst-case interior-point implementation may scale as $O(K^6)$; however, this represents a conservative upper bound. In practice, sparsity and problem structure significantly reduce complexity.

The dominant runtime is typically contributed by local training or server-side optimization depending on system scale. Overall, the proposed framework remains suitable for moderate-scale edge deployment with centralized coordination.

\section{Conclusion}

This paper proposes a closed-loop personalized OTA-FL framework for time-varying RIS-assisted cascaded channel estimation. The framework integrates GRU-based temporal learning, personalized federated optimization, over-the-air model aggregation, RIS-enabled physical-layer control, and dynamics-aware grouping into a unified communication and learning system. The GRU-based estimator leverages temporal correlations across communication rounds to jointly estimate current CSI and predict uplink-instant channels. This design reduces pilot and uplink mismatch under mobility. The shared-backbone architecture with task-specific local heads enables personalization while maintaining compatibility with over-the-air aggregation under non-IID data distributions. To improve robustness in heterogeneous environments, a warmup-then-group strategy is introduced. The system first performs global aggregation and then switches to groupwise collaboration once stable representations are formed. Theoretical analysis shows convergence of the Reptile-based shared backbone under OTA distortion. It also establishes stationary-point convergence for the SCA-based grouping algorithm. These results clarify how channel estimation errors, communication distortion, and data heterogeneity jointly influence learning dynamics. Numerical results demonstrate consistent improvements in uplink channel prediction accuracy and OTA aggregation performance. The gains are especially evident in low-pilot regimes, high mobility scenarios, and settings with strong tail-user effects. Overall, the results indicate that temporal channel learning, personalized federated optimization, physical-layer design, and dynamics-aware grouping are mutually supportive components. Their joint optimization is essential for reliable distributed learning in dynamic wireless environments.

\appendices

\section{Proof of Lemma 1}
\label{app:lemma1}

Let \(S^{r}=\sum_{k=1}^{K}\omega_{k}^{r}\) and
\(\vartheta_{k}^{r}\triangleq c_{k}^{r}(\hat c_{k}^{r})^{*}/|\hat c_{k}^{r}|^{2}-1\). With
\(p_{k}^{r}=\sqrt{\eta^{r}}\omega_{k}^{r}\nu_{k}^{r}(\hat c_{k}^{r})^{*}/|\hat c_{k}^{r}|^{2}\), the component-wise error is
\begin{align}
\hat r_{\boldsymbol\phi}^{r}[d]-r_{\boldsymbol\phi}^{r}[d]
&=
\frac{1}{\sqrt{\eta^{r}}}\sum_{k=1}^{K}c_{k}^{r}p_{k}^{r}s_{k}^{r}[d]
+\frac{n^{r}[d]}{\sqrt{\eta^{r}}}
-\sum_{k=1}^{K}\omega_{k}^{r}\nu_{k}^{r}s_{k}^{r}[d] \notag\\
&=
\sum_{k=1}^{K}\omega_{k}^{r}\nu_{k}^{r}\vartheta_{k}^{r}s_{k}^{r}[d]
+\frac{n^{r}[d]}{\sqrt{\eta^{r}}},
\end{align}
and from \eqref{eq:ota_e}, we obtain
\begin{equation}
\mathbf e_{\mathrm{OTA}}^{r}
=
\sum_{k=1}^{K}\omega_{k}^{r}\nu_{k}^{r}\vartheta_{k}^{r}\mathbf s_{k}^{r}
+
\frac{\mathbf n^{r}}{\sqrt{\eta^{r}}},
\quad
\mathbf n^{r}\sim\mathcal{CN}(\mathbf 0,\sigma_n^{2}\mathbf I_{D_{\boldsymbol\phi}}).
\end{equation}
Conditioned on \(\mathcal H_r\), \(\mathbf n^r\) is independent and zero-mean, so, using
\(\|\mathbf s_k^r\|_2^2=D_{\boldsymbol\phi}\), Cauchy--Schwarz and
\((\omega_k^r\nu_k^r)^2\le P_t|\hat c_k^r|^2/\eta^r\) from \eqref{eq:eta_r},
\begin{align}
\mathbb E
\left[
\|\mathbf e_{\mathrm{OTA}}^{r}\|_{2}^{2}
\mid\mathcal H_r
\right]
&=
\left\|
\sum_{k=1}^{K}\omega_{k}^{r}\nu_{k}^{r}\vartheta_{k}^{r}\mathbf s_{k}^{r}
\right\|_{2}^{2}
+
\frac{D_{\boldsymbol\phi}\sigma_n^{2}}{\eta^{r}} \notag\\
&\le
D_{\boldsymbol\phi}
\left(
\sum_{k=1}^{K}\omega_{k}^{r}\nu_{k}^{r}|\vartheta_{k}^{r}|
\right)^2
+
\frac{D_{\boldsymbol\phi}\sigma_n^{2}}{\eta^{r}} \notag\\
&\le
\frac{D_{\boldsymbol\phi}\sigma_n^{2}}{\eta^{r}}
\left(
1+\Xi_{\mathrm{CSI}}^{r}
\right).
\end{align}
Substituting
\(
1/\eta^r
=
[(S^r)^2]\mathrm{NMSE}_{\mathrm{proxy}}^r/(\sigma_n^2P_t)
\) gives
\begin{align}
\mathbb E
\left[
\|\mathbf e_{\mathrm{OTA}}^{r}\|_{2}^{2}
\mid\mathcal H_r
\right]
&\le
D_{\boldsymbol\phi}
\frac{(S^r)^2}{P_t}
\left(
1+\Xi_{\mathrm{CSI}}^{r}
\right)
\mathrm{NMSE}_{\mathrm{proxy}}^r \notag\\
&\triangleq
C_r\mathrm{NMSE}_{\mathrm{proxy}}^r .
\end{align}
If \(\sum_k|\hat c_k^r|^2\le C_{\hat c}\), \(\sum_k|\vartheta_k^r|^2\le C_\gamma\), \(0<\underline S\le S^r\le K\bar\omega\), then
\begin{equation}
C_r
\le
\overline C
\triangleq
D_{\boldsymbol\phi}
\frac{(K\bar\omega)^2}{P_t}
\left(
1+\frac{P_t}{\sigma_n^2}C_{\hat c}C_\gamma
\right),
\end{equation}
which gives \eqref{eq:lemma1_uni}.

\section{Proof of Theorem 1}
\label{app:theorem1}

By definition, $\Delta\boldsymbol\phi^r=\beta_R\hat{\mathbf r}_{\boldsymbol\phi}^r=\beta_R\mathbf r_{\boldsymbol\phi}^r+\beta_R\mathbf e_{\mathrm{OTA}}^r$. Under \eqref{eq:descent_lemma_phi}, and with $\mathbb E[\mathbf e_{\mathrm{OTA}}^r\mid\boldsymbol\phi^r]=\mathbf 0$, $\mathbb E[\mathbf r_{\boldsymbol\phi}^r\mid\boldsymbol\phi^r]
=-\nabla F(\boldsymbol\phi^r)+\mathbf b^r$ (where $\mathbf b^r$ denotes the non-IID personalization bias and satisfies \(\|\mathbf b^r\|_2^2\le B_{\mathrm{het}}^r\)), and Young's inequality, we have
\begin{align}
\mathbb E[
F(\boldsymbol\phi^{r+1})\mid\boldsymbol\phi^r]
&\le
F(\boldsymbol\phi^r)
-\beta_R
\left(
1-\frac{\kappa}{2}
\right)
\|\nabla F(\boldsymbol\phi^r)\|_2^2 \notag\\
&
+\frac{\beta_R}{2\kappa}B_{\mathrm{het}}^r
+\frac{L}{2}
\mathbb E[
\|\Delta\boldsymbol\phi^r\|_2^2
\mid
\boldsymbol\phi^r].
\label{eq:Lsmooth_bound}
\end{align}
Moreover, using $\|\Delta\boldsymbol\phi^r\|_2^2
\le
2\beta_R^2\|\mathbf r_{\boldsymbol\phi}^r\|_2^2
+
2\beta_R^2\|\mathbf e_{\mathrm{OTA}}^r\|_2^2$, $\mathbb E[
\|\mathbf r_{\boldsymbol\phi}^r\|_2^2
\mid
\boldsymbol\phi^r]
\le
G_1\|\nabla F(\boldsymbol\phi^r)\|_2^2
+
\sigma_{\mathrm{loc}}^{2,r}$ and \eqref{eq:lemma1_uni}, which gives
\begin{align}
\frac{L}{2}
\mathbb E[
\|\Delta\boldsymbol\phi^r\|_2^2
\mid
\boldsymbol\phi^r]
&\le
L\beta_R^2G_1
\|\nabla F(\boldsymbol\phi^r)\|_2^2 \notag\\
& \quad
+
L\beta_R^2\sigma_{\mathrm{loc}}^{2,r}
+
L\beta_R^2\overline C
\mathrm{NMSE}_{\mathrm{proxy}}^r .
\label{eq:app_delta_bound}
\end{align}
Combining \eqref{eq:Lsmooth_bound} and \eqref{eq:app_delta_bound}, define
$c_1\triangleq\beta_R(1-\frac{\kappa}{2})-L\beta_R^2G_1$,
$c_2\triangleq\frac{1}{2\kappa}$,
$c_3\triangleq L$, and
$c_4\triangleq L\overline C$.
For $\kappa\in(0,2)$, choosing
$0<\beta_R<\frac{1-\kappa/2}{LG_1}$
ensures $c_1>0$, making
$-c_1\|\nabla F(\boldsymbol\phi^r)\|_2^2$
a descent term. We thus obtain
\begin{equation}
\begin{aligned}
\mathbb E[
F(\boldsymbol\phi^{r+1})
\mid
\boldsymbol\phi^r]
&\le
F(\boldsymbol\phi^r)
-c_1\|\nabla F(\boldsymbol\phi^r)\|_2^2
+c_2\beta_RB_{\mathrm{het}}^r \\
&\quad +c_3\beta_R^2\sigma_{\mathrm{loc}}^{2,r}
+c_4\beta_R^2\mathrm{NMSE}_{\mathrm{proxy}}^r .
\end{aligned}
\label{eq:app_one_step}
\end{equation}
Taking total expectation, summing \eqref{eq:app_one_step} over \(r=0,\ldots,R-1\), and using \(F\ge F_{\inf}\),
\begin{gather}
\frac{1}{R}
\sum_{r=0}^{R-1}
\mathbb E
\|\nabla F(\boldsymbol\phi^r)\|_2^2
\le
\frac{\mathbb E[F(\boldsymbol\phi^0)]-F_{\inf}}{c_1R}
+
\frac{c_2\beta_R}{c_1R}
\sum_{r=0}^{R-1}
B_{\mathrm{het}}^r\\
+
\frac{c_3\beta_R^2}{c_1R}
\sum_{r=0}^{R-1}
\sigma_{\mathrm{loc}}^{2,r} \notag
+
\frac{c_4\beta_R^2}{c_1R}
\sum_{r=0}^{R-1}
\mathrm{NMSE}_{\mathrm{proxy}}^r .
\end{gather}
The claimed stationarity bound follows. As $R\to\infty$, the $\mathcal{O}(1/R)$ initial-gap term vanishes, leaving a stationary neighborhood determined by averaged heterogeneity, local stochasticity, and OTA distortion; if these disturbance averages vanish, the time-averaged squared gradient norm converges to zero. Since $\mathrm{NMSE}_{\mathrm{proxy}}^r$ decreases with increasing $\eta^r$, improving worst-user reliability reduces the OTA-induced steady-state error.

\bibliographystyle{IEEEtran}
\bibliography{ieeetran}

\end{document}